\documentclass[11pt]{article}

\usepackage{natbib}
\usepackage{graphicx}

\newcounter{tr}
\newcommand{\bqe}{\begin{eqnarray}}
\newcommand{\eqe}{\end{eqnarray}}

\begin{document}

\title{Theory of dynamic crack branching in brittle materials}
\author{E. Katzav$^\dag$\footnote{Corresponding author. Email: eytan.katzav@lpt.ens.fr} , M. Adda-Bedia$^\dag$ and R. Arias$^\ddag$\\
$^\dag$Laboratoire de Physique Statistique de l'Ecole Normale Sup\'erieure, CNRS UMR 8550,\\
24 rue Lhomond, 75231 Paris Cedex 05, France.\\
$^\ddag$Departamento de F\'{\i}sica, Universidad de Chile, Casilla 487-3, Santiago, Chile.}
\date{\today}

\maketitle

\begin{abstract}

The problem of dynamic symmetric branching of a tensile crack
propagating in a brittle material is studied within Linear Elastic
Fracture Mechanics theory. The Griffith energy criterion and the
principle of local symmetry provide necessary conditions for the
onset of dynamic branching instability and for the subsequent paths
of the branches. The theory predicts a critical velocity for
branching and a well defined shape parameterized by a branching angle and
a curvature of the side branches. The model rests on a scenario of
crack branching based on reasonable assumptions and on exact dynamic
results for the anti-plane branching problem. Our results reproduce
within a simplified $2D$ continuum mechanics approach the main
experimental features of the branching instability of fast cracks in
brittle materials.

\end{abstract}

\section{Introduction}
\label{section:intr}

The continuum theory of fracture mechanics is concerned with the quantitative description of the mechanisms of crack nucleation, the conditions under which they propagate and their dynamics \citep{Freund,Broberg}. For brittle materials, the relationship between internal stress and deformation and the balance laws of physics dealing with mechanical quantities do not include the possibility of material separation. Consequently, the equation of motion of the crack front is based on additional statements on crack growth. The most frequently used criterion of crack propagation in two dimensional elastic brittle materials consists of two parts: Griffith's hypothesis and the principle of local symmetry.

The Griffith's energy criterion \citep{Griffith20,Freund,Broberg} states that the intensity of the loading necessary to produce propagation is given by ${\cal G}=\Gamma$, where ${\cal G}$ is the energy release rate and $\Gamma$ is the fracture energy of the material (i.e. the energy needed to create new surfaces). The principle of local symmetry states that the crack advances in such a way that in-plane shear stresses always vanish in the vicinity of the crack tip. This rule was first proposed for quasi-static cracks \citep{Goldstein,Leblond89}, and generalized to rapidly moving cracks \citep{adda99}. It was shown in \citep{adda99} that the two criteria arise from the same physical origin. The energy release rate is the component of the driving force along the direction of crack motion, $F_{1}$. Griffith's energy criterion may then be reinterpreted as a material force balance between $F_{1}$ and a force that resists the crack advance, i.e. $F_{1} = \Gamma$. However, this equation is not sufficient to determine the trajectory of a crack. If one assumes that material force balance holds at the crack tip, one should impose the component of the material force perpendicular to the direction of crack propagation to vanish. This condition is identically satisfied if the loading in the vicinity of the crack tip is purely tensile.

The Griffith criterion and the principle of local symmetry predict adequately the path and the stability of slowly propagating cracks \citep{adda95,adda96,Bouch03,Marder04}. Controlled experiments on quasi-static cracks confirm the theoretical results \citep{Yuse93,Ronsin95}. In the case of fast crack propagation, experiments on different brittle materials \citep{Ravi84,Fineberg92,Gross93,Boudet96,Sharon95,Sharon96,Sharon99,Fineberg05} have identified a dynamic instability related to a transition from a single crack to a branched crack configuration. The instability occurs when the crack speed exceeds a critical velocity $v_c$, which does not depend on the applied traction and on the geometry of the plate. Above $v_c$, a single crack is no longer stable. Instead, a repetitive process of micro-branching occurs, which changes the crack dynamics~: the acoustic emission from the crack increases \citep{Boudet96,Boudet00,Gross93}, the crack velocity develops strong oscillations and a pattern, which is correlated with the velocity oscillations, is observed on the fracture surface \citep{Fineberg92,Sharon95,Sharon96, Sharon99,Fineberg05}.

Some aspects of this dynamic instability were described in the framework of the theory of brittle fracture mechanics \citep{adda04,adda05,Bouch05}. These studies were based on Eshelby's approach which states that, as in the single crack case, a growth criterion for a branched crack must be based on the equality between the energy flux into the two propagating tips and the energy required to open the material and create new surfaces as a result of this propagation \citep{Eshelby70}. The problem of determining the in-plane dynamic stress intensity factors immediately after branching was formulated in \citep{adda05}. It was shown that the in-plane elastic fields immediately after branching exhibit self-similar properties, and that the corresponding stress intensity
factors do not explicitly depend on the velocity of the single crack tip before branching. These properties are similar to the mode~III crack branching problem, which was solved exactly in \citep{adda03,adda04b}. This similarity suggests that under plane loading configurations, the jump in the energy release rate due to branching is maximized when the branches start to propagate very slowly. Under this assumption, the branching of a single propagating crack under tensile loading was found to be energetically possible when its speed exceeds a certain critical value \citep{adda05}. The theoretical results for the critical velocity and the branching angle agree fairly well with the available experimental results \citep{Fineberg92, Sharon96, Sharon99,Fineberg05}.

The main purpose of the present study is to perform a quantitative analysis of the subsequent paths followed by the branches. Following \citep{Karihaloo81,Leblond89,amle}, the asymptotic expansion of the stress field at the tip of a curved extension of a branched crack is presented. Using these exact results, the paths selected by the branched cracks are derived. As a main result, the experimentally observed shape of the branches \citep{Sharon96, Sharon99,Fineberg05} is recovered without introducing any additional parameters. The present study shows that both the branching instability threshold, the branching angle and the subsequent paths of the branches can be predicted within the continuum $2D$ theory of brittle fracture mechanics. Note that the present analysis provides a necessary condition for branching and not an instability mechanism for it. In addition, the branching instability in real systems is of $3D$ nature \citep{livne06}. Therefore, the present study should be seen as a step towards a complete understanding of this phenomenon.

For the sake of clearness, the second section of this paper summarizes the results obtained for the branching problem: the branches shape, their dynamics as well as the dynamic instability are determined.
In this paper a systematic analysis of the branching problem is made, and due to reasons of completeness, some of the results already communicated in \citep{adda04,adda05} are presented here again. Higher order terms than those in \citep{adda05} were calculated, as well as numerical corrections of lower order terms are given. The third section of the paper presents the  detailed study of the branching problem. The fourth and last section solves an elastostatic problem related to the experimental setups \citep{Fineberg92,Sharon95,Sharon96, Sharon99,Fineberg05}, where the interest is to determine possible outcomes of the non singular $T$ stress at the original crack tip.

\section{Summary of results and future prospects}
\label{section:sumofres}
First, a static analysis of crack branching under plane loading conditions is done. It is an exact approach that follows that of \citep{amle} for the kinking case. Using the principle of local symmetry as a criterion for determining crack's trajectories, as well as asymptotic analysis of the local stress configurations at the crack tips of branches of a given longitude and characteristic shape, the crack's branching angle and subsequent curved paths are determined. This is done by solving integral equations for the elastic potentials, with the help of conformal mapping and perturbation techniques.

Secondly, the dynamic crack branching  is addressed by arguing that the plane loading case should not differ very much qualitatively from the exact solutions of anti-plane branching, where indeed dynamic branching results can be safely retrieved from the static ones. Much of this is based on the argument that it is plausible that the branching occurs with the new branches starting at vanishing speeds since the elastic energy release
rate at each crack tip is maximal in that case. Under these assumptions, application of Griffith's energy criterion leads to the critical speed for branching (once all the post-branching elastic quantities are replaced by their static counterparts).

These results, detailed in the following sections, support the following scenario for the process of dynamic branching instability in brittle materials

\begin{itemize}
\item The critical velocity at which the crack tip can branch depends on the material parameters only through the fracture energy and the elastic constants. This prediction for the critical velocity agrees with the available experimental results.
\item The paths that the branches take consists of two universal features: a branching angle of $27^o$
followed by a curved extension described by a single curvature parameter which is calculated below. The branching angle as well as the general shape of the extensions coincide with those seen in the experiments.
\item The velocity of the branches is vanishingly small right after branching, which seems to be a peculiar characteristic. However, since the predicted acceleration of the tips after branching is very high, this may explain the experimental results.
\end{itemize}

The above scenario for a single branching event can now be integrated into a general picture of the propagation dynamics of the crack,
which can reproduce the fractography of the broken surface, by considering in addition the stability analysis of the branched configuration reported in \citep{Bouch05}.
It turns out that the symmetric form of branching is unstable, in the sense that sooner or later one extension continues to grow while the other one slows down until it stops.
This leads to a pattern of a broken surface composed of a sequence of branching events, where each time only one extension survives.
The surviving crack tip accelerates between branching events and then decelerates abruptly in the next branching event.
Although the $3D$ nature of the instability is not taken into account here \citep{livne06}, we think that this scenario gives a coherent physical interpretation of the fractography of broken samples. In order to describe completely the observed patterns, our $2D$ analysis should be coupled to an instability mechanism of the crack front itself. However, we expect that features such as the critical velocity for branching or the branches' shape should be modified by the $3D$ nature of the problem as a secondary effect.

In the following section we detail the results that led to the above mentioned conclusions.

\subsection{Stress field ahead of curved extensions of a branched crack}

\begin{figure}[ht]
\centerline{\includegraphics[width=8cm]{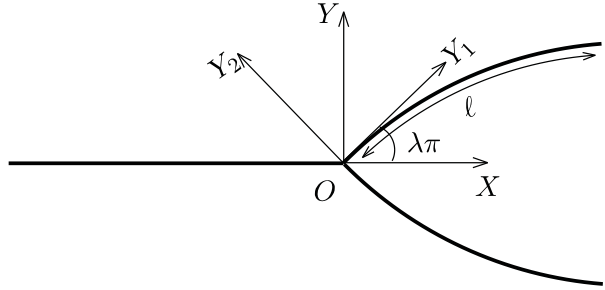}}
\caption{Schematic representation of a straight crack with two symmetrically branched curved extensions.}
\label{fig:problem}
\end{figure}

Consider an elastic body containing a straight crack with symmetrically branched curved extensions of length $\ell$ and a branching angle $\lambda\pi$ (see Fig.~\ref{fig:problem}). Let  $XOY$ denote the coordinate system with the $OX$ axis directed along the initial straight crack, and let $Y_1OY_2$ denote the coordinate system with the $OY_1$ axis directed along the tangent to the upper extension at the point $O$. These two coordinate systems are obviously related by
\begin{eqnarray}
\label{eq:rotate}
Y_1 & = & X\cos\lambda\pi+ Y\sin\lambda\pi\;,\\
Y_2 & = & Y\cos\lambda\pi- X\sin\lambda\pi \;.
\end{eqnarray}
Following the approach developed in \citep{Karihaloo81,Leblond89,adda04}, it can be shown that the
asymptotic shape of the crack extension is necessarily given by
\begin{equation}
\label{eq:shape}
Y_2=aY_1^{3/2}+O(Y_1^2)\;,
\end{equation}
where $a$ is a curvature parameter whose dimension is $(\mbox{length})^{-1/2}$. Moreover, the expansion of the static stress intensity factors $K^{\prime}_l(s)$ $(l=1,2)$ at the crack tip in powers of $\ell$ obeys the general form \citep{Leblond89}
\begin{eqnarray}
K'_l(\ell) &=& K_l^*+K_l^{(1/2)}\sqrt{\ell} + O(\ell) \nonumber \\
&=& \sum_{m=1,2}F_{lm}(\lambda) K_m + \sum_{m=1,2}
\left[G_m(\lambda) T \delta_{lm} + a H_{lm}(\lambda) K_m \right]
\sqrt{\ell} + O(\ell) \;.\label{eq:SIFs}
\end{eqnarray}
In this expansion, $K_l$ and $T$ are the static stress intensity factors and the nonsingular stress
in the universal expansion of the stress field at the original crack tip $O$ without the branched
extensions. $K_l$ and $T$ are given by
\begin{equation}
\label{eq:exp0}
\sigma_{ij}(r,\theta)=\sum_{l=1,2}\frac{K_l}{\sqrt{2\pi r}}\,\Sigma^{(l)}_{ij}(\theta)
+T\delta_{iX}\delta_{jX}+O\left(\sqrt{r}\right)\;,
\end{equation}
where $(r,\theta)$ are polar coordinates referred  to the branching point $O$, and $\Sigma^{(l)}_{ij}$ are known functions describing the angular variations of the stress field components \citep{Broberg}. The functions $F_{lm}$, $G_l$ and $H_{lm}$ are universal in the sense that they do not depend neither on the geometry of the body nor on the applied loading. They depend only  on the branching angle and their computation can be performed following the approach developed in \citep{amle}. Note that the asymptotic expansions given by Eqs.~(\ref{eq:shape})-(\ref{eq:SIFs}) are applicable to crack extensions obtained by actual propagation of the initial crack and not simply by arbitrary machining of the body \citep{Leblond89}. Due to the linearity of the problem, the expressions (\ref{eq:shape})-(\ref{eq:SIFs}) can be predicted from dimensional arguments. Since the $K_l$'s scale as stress$\times\sqrt{\mbox{length}}$ and $T$ scales as stress, the first order expansion of the stress intensity factors in (\ref{eq:SIFs}) must involve an additional parameter whose dimension is $1/\sqrt{\mbox{length}}$. This parameter is provided by the asymptotic expansion (\ref{eq:shape}) of the branched extension.

\begin{figure}[ht]
\centerline{\includegraphics[width=16cm]{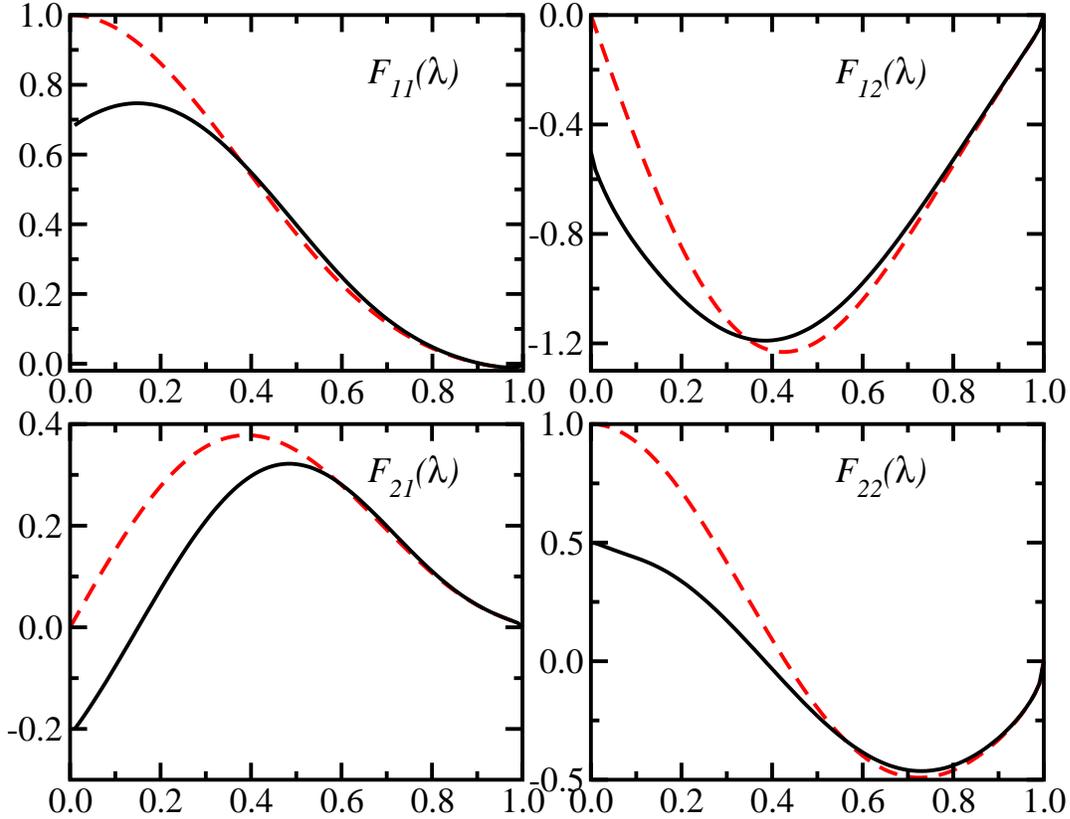}}
\caption{The elements of the matrix $F_{lm}(\lambda)$ for
the symmetrically branched crack (solid lines) and for the kinked
crack (dashed lines).} \label{fig:Flm}
\end{figure}
\begin{figure}[ht]
\centerline{\includegraphics[width=16cm]{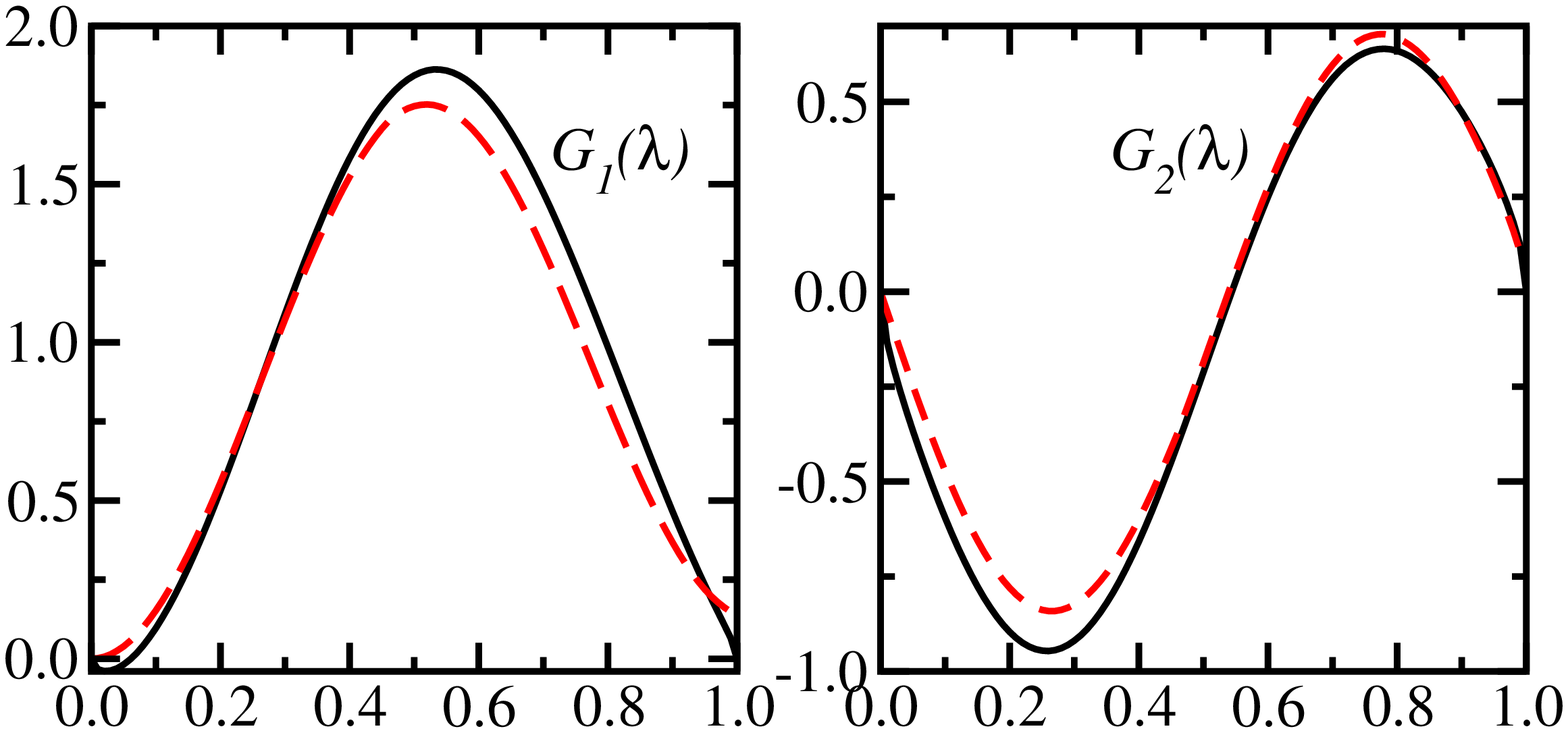}}
\caption{The elements of the vector $G_l(\lambda)$ for the
symmetrically branched crack (solid lines) and for the kinked crack
(dashed lines).} \label{fig:Gl}
\end{figure}
\begin{figure}[ht]
\centerline{\includegraphics[width=16cm]{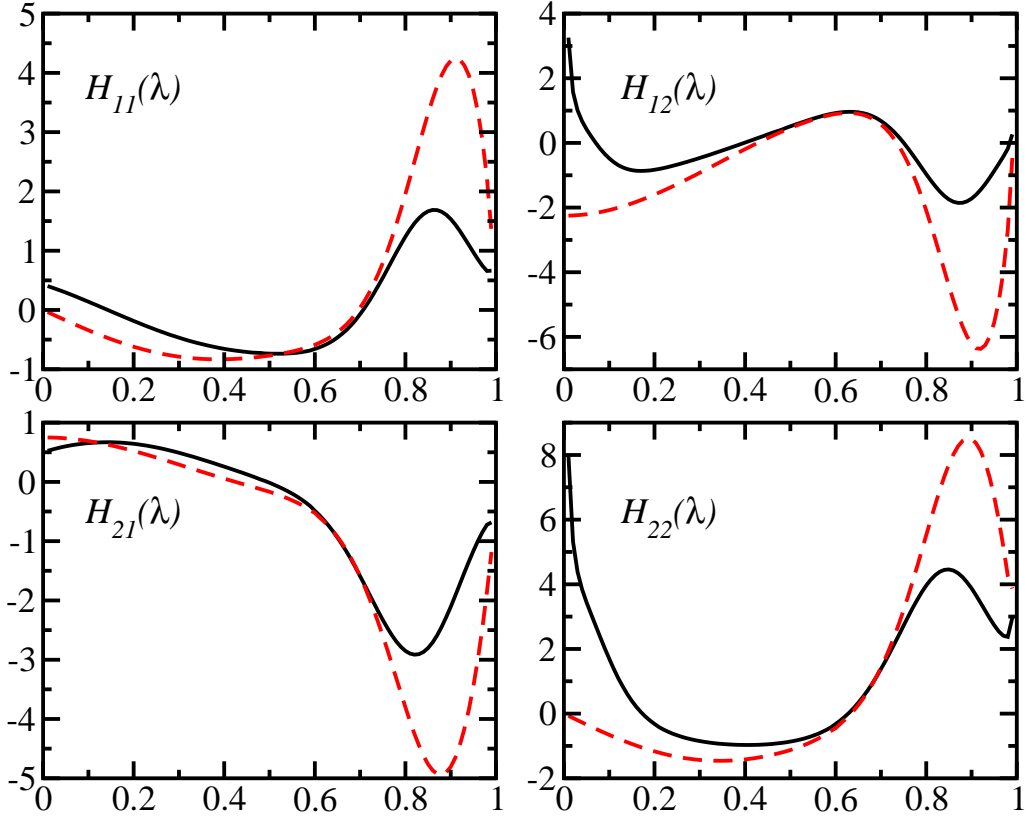}}
\caption{The elements of the matrix $H_{lm}(\lambda)$ for
the symmetrically branched crack (solid lines) and for the kinked
crack (dashed lines).} \label{fig:Hlm}
\end{figure}

In Figures~\ref{fig:Flm}, \ref{fig:Gl} and \ref{fig:Hlm} we present the results for the universal functions $F_{lm}$, $G_l$ and $H_{lm}$ for the branched cracks. As a comparison, we also present the results obtained for these functions for the kinked crack problem \citep{amle}. The functions $F_{lm}$ for a symmetrically branched configuration have been already computed in \citep{adda05}, while the computation of $G_l$ and $H_{lm}$ is new.  Details of this analysis are given in the following section. Note that the results for the functions $F_{lm}$ in Fig.~\ref{fig:Flm} correct numerical inaccuracies in the results reported in \citep{adda04}. These inaccuracies have of course a quantitative implication by correcting certain values such as the critical branching angle. However, more importantly, it contradicts the statement made in \citep{adda04} concerning a possible difference between the predictions of the principle of local symmetry and the maximal energy release rate criterion, as will be explained below.

Once a detailed expansion of the stress intensity factors is available, it remains to be combined with a propagation criterion, in order to get crack path prediction. Griffith's energy criterion \citep{Griffith20,Freund,Broberg}
and the principle of local symmetry \citep{Goldstein,Leblond89} imply
\begin{eqnarray}
{\cal G}^{\prime}_1(s) = \frac{1}{2\mu}K^{\prime2}_1(s) & = & \Gamma \;,
\label{eq:grif}\\
    K^{\prime}_2(s) & = & 0 \;,
\label{eq:pls}
\end{eqnarray}
where $\mu$ is the Lam\'e shear coefficient of the material. Note that Eq.~(\ref{eq:pls}) imposes the symmetry of the stress field in the vicinity of the crack tip which in turn affects the crack direction of propagation. Therefore, the crack path is mainly selected by the principle of local symmetry, while Eq.~(\ref{eq:grif}) controls the intensity of the loading necessary to advance the crack. In the following, the stability of a tensile crack and the path selection of branched cracks will be discussed in
view of these general results.

\subsection{The shape of the branched cracks}

\begin{figure}[ht]
\centerline{\includegraphics[width=8cm]{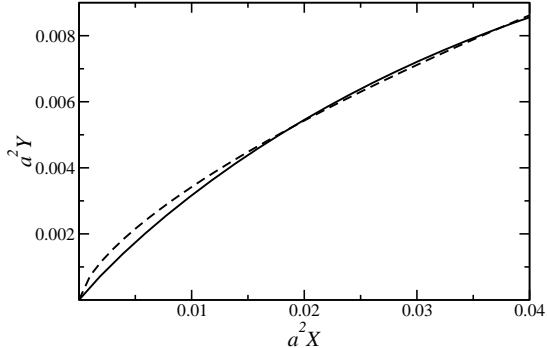}}
\caption{Path followed by the branch when $a<0$, and a comparison with the functional form $Y(X)\sim X^{2/3}$ proposed in \citep{Sharon99}.}
\label{fig:shape}
\end{figure}

Consider a straight crack initially under pure tensile (mode~I) loading that branches into two symmetrical cracks. As shown in \citep{adda04}, the paths of the branches can be determined if each crack tip satisfies the principle of local symmetry during propagation \citep{Goldstein}. Imposing this (Eq.~(\ref{eq:pls})), the following two conditions that determine the branching angle $\lambda$ and the curvature parameter $a$ should be satisfied
\begin{eqnarray}
F_{21}(\lambda)=0  \label{eq:lambdaCR}, \\
a = - \frac{G_2(\lambda)}{H_{21}(\lambda)} \frac{T}{K_1}
\label{eq:a}.
\end{eqnarray}
Using the results presented in Figures~\ref{fig:Flm}, \ref{fig:Gl} and \ref{fig:Hlm}, Eq.~(\ref{eq:lambdaCR}) gives $\lambda=\lambda_c=0.15$ corresponding to a branching angle of $27^o$, in agreement with the results of \citep{Isida92,adda05}. Also we find that $G_2(\lambda_c)/H_{21}(\lambda_c) = -1.16$, and consequently the sign of $a$ is determined by the sign of the ratio $T/K_1$. Therefore, the convexity of the branches' paths depends on the sign of the $T$-stress: If $T<0$ the branches will tend to come back towards the initial direction of the crack prior to branching (see Fig.~\ref{fig:shape}), while if $T>0$ the branches will diverge away from this direction.

Before continuing, it is useful to define the material parameter $\kappa=(c_d/c_s)^2$, where $c_d$ and $c_s$ are the dilatational and shear wave speeds. This quantity will serve to compare theoretical results with experiments. $\kappa$ is related to the Poisson ratio $\nu$ by \citep{Broberg}
\begin{equation}
\left( {\frac{{c_d }}{{c_s }}} \right)^2  \equiv \kappa  = \left\{
\begin{array}{l}
 \frac{2}{{1 - \nu }}\quad {\mbox{for plane stress}} \\
 \frac{{2 - 2\nu }}{{1 - 2\nu }}\quad {\mbox{for plane strain.}} \\
 \end{array} \right.
\label{eq:kappa}
\end{equation}

In order to estimate the $T$-stress for the real experimental setups of \citep{Fineberg92,Sharon99,Fineberg05}, we solved the elastostatic problem of a semi-infinite straight crack in an infinite strip of half-width $W$, whose top and bottom edges are fixed at positions $\pm\delta$ above their initial positions (see Fig. \ref{fig:Tstresss}). The analysis has been performed for both clamped edges and shear free edges boundary conditions. The details of the calculations are given in section \ref{section:T}. It turns out that for typical situations the ratio $T/K_1$ is negative and so the branches will be stable, i.e. they will not diverge away from the initial direction of the primary crack (see Fig.~\ref{fig:TPr1} and Fig.~\ref{fig:TPr2}a). For the typical value of $\kappa=3$, we get $T/K_1=-0.43/\sqrt{W}$ when the edges of the strip are clamped, which leads to a value of the curvature parameter $a = -0.5/\sqrt W$. This theoretical value is consistent with the estimation of the parameter $a$ from the experimental results of \citep{Sharon96,Sharon99}. However, a more quantitative comparison with the experiments would require to take into account finite size effects, the three dimensional geometry of the sample and dynamical effects.

It is interesting to mention that when the edges are shear free, the sign of the $T$-stress can be made positive by applying an additional loading $T_{\infty}$ by stretching the strip in the $X$-direction. This results in the existence of a critical $T_{cr}(\kappa)$ such that the path of the branches becomes unstable, i.e. diverges away from the direction of the primary crack (see Fig.~\ref{fig:TPr2}b). Such a situation where different boundary conditions can change the stability of the propagating crack is of interest, and can certainly be tested experimentally.

\subsection{The dynamic branching instability}

Until now, only static aspects of the branching instability have been discussed. In order to address dynamical aspects of this problem, such as determining the onset of dynamic branching, let us consider the following scenario: A semi infinite straight crack that propagates at a speed $v(t)$ for $t<0$ suddenly stops at $t\rightarrow 0^-$. At $t\rightarrow 0^+$, the crack branches locally with a branching angle equal to $\lambda\pi$. For $t>0$, the new branches propagate at a velocity $v'(t)$ in the new directions $\pm\lambda\pi$. It is well established \citep{Kostrov75,Freund} that the dynamic stress intensity factors, $K_l(t,v)$, of the straight crack prior to branching are related to the rest stress intensity
factors, $K_l(t,0)$, of the same configuration by
\begin{equation}
K_l(t,v)=k_l(v)K_l(t,0),
\label{eq:dynamicSIF1}
\end{equation}
where $k_l(v)$ ($l=1,2,3$) are known universal functions of the instantaneous crack tip speed $v(t)$, whose explicit forms can be found in \citep{Freund,Broberg}.

Since in this problem there is no time scale, and consequently no length scale, against which the independent variables can be scaled, the dynamic stress intensity factors immediately after branching, $K'_l(0^+,v',v)$ can always be written in the form of a universal function of the velocities and branching angle, multiplied by the static stress intensity factors immediately before branching, $K_l(0^-,v=0)$, i.e.
\begin{equation}
K'_l(0^+,v',v)=\sum_m k_l(v') \mathcal{F}_{lm}(\lambda,v',v) K_l(0^-,0).
\label{eq:dynamicSIF2}
\end{equation}
As in the quasi-static case \citep{Leblond89}, the matrix
$\mathcal{F}$ is universal in the sense that it depends neither on
loading configuration nor on the geometry of the body. Indeed in
the limit that is
considered here, the dynamic branching problem does not involve
radiation effects, so it is always equivalent to a crack
propagating in an unbounded body. Moreover, $\mathcal{F}$ should
approach the elastostatic solution for a vanishingly small
velocity of the side-branches, namely
\begin{equation}
\lim_{v'\rightarrow0} \mathcal{F}_{lm}(\lambda,v',v)=
F_{lm}(\lambda)\;.
\label{eq:dynamicSIF3}
\end{equation}
Based on the solution to the anti-plane branching problem \citep{adda03,adda04b}, it was shown in \citep{adda05} that the dependence of the velocity of the single crack tip before branching is suppressed from the stress distribution that has to be balanced during the propagation of the branches. Consequently,
the matrix elements $\mathcal{F}_{lm}$ related to plane loading situations should also be independent of the velocity prior to branching, namely
\begin{equation}
\mathcal{F}_{lm}(\lambda,v',v)= \mathcal{F}_{lm}(\lambda,v'),
\label{eq:dynamicSIF4}
\end{equation}

\begin{figure}[ht]
\centerline{\includegraphics[width=8cm]{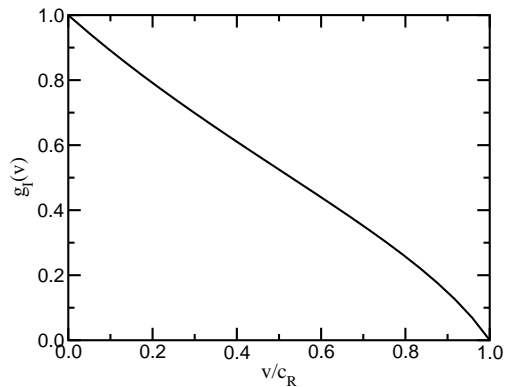}}
\caption{The universal function $g_1(v)$ for $\kappa= 3$. Here and elsewhere, $c_R$ denotes the Rayleigh wave speed.}
\label{fig:Gdynamic}
\end{figure}

In order to proceed, one must come up with a growth criterion for a branched crack. It is well established that the dynamic energy release rate $G$ for a single straight crack is given by
\citep{Kostrov75,Freund}
\begin{equation}
G=\frac{1}{2\mu}\sum_{l=1}^3A_l(v)K_l^2(t,v)=\frac{1}{2\mu}\sum_{l=1}^3g_l(v)K_l^2(t,0)\;,
\label{eq:Griffith1}
\end{equation}
where
\begin{equation}
g_l(v)=A_l(v)k_l^2(v)\;.
\end{equation}
The functions $A_l(v)$ and $g_l(v)$ do not depend on the details of the applied loading,  nor on the configuration of the body being analyzed. They only depend on the local instantaneous speed of the
crack tip and on the properties of the material \citep{Freund}. Fig.~\ref{fig:Gdynamic} shows the function $g_1(v)$ that will be used in the following.

A growth criterion for a branched crack must also be based on the equality between the elastic energy flux into each propagating tip and the energy that is used in creating new broken surface during this propagation \citep{Griffith20}. The dynamic energy release rate is a quantity associated to a single moving crack tip, and so it has to be determined for each crack tip. When the primary crack before branching is under pure mode~I loading and due to the symmetry of the branching configuration, the energy release rate immediately after branching $G'$ for each crack tip is given by
\begin{equation}
G'=\frac{1}{2\mu}\left[g_1(v')\mathcal{F}_{11}^2(\lambda,v')
+g_2(v')\mathcal{F}_{21}^2(\lambda,v') \right]K_1^2(0^-,0).
\label{eq:Griffith2}
\end{equation}

According to the generalized Griffith's criterion \citep{Griffith20}, the crack must grow in such a way that the energy release rate is always equal to the dynamic fracture energy of the material, $\Gamma(v)$, which is assumed to be a property of the material and whose value may depend on the instantaneous crack tip speed \citep{Freund,Boudet96,Sharon99}. This growth criterion should hold for the crack tips before and after branching, and so it introduces a relation between the energy release rates immediately before and after branching, namely
\begin{equation}
G'=\frac{\Gamma(v')}{\Gamma(v)} G,
\label{eq:Griffith3}
\end{equation}
which is a {\em nec essary} condition for the existence of a branching configuration. If this condition is not fulfilled then single crack tip propagation would be maintained. Let us stress again that this condition does not provide an instability mechanism to the branching process. However, whichever instability mechanism, it should respect Eq.~(\ref{eq:Griffith3}).

\begin{figure}[ht]
\centerline{\includegraphics[width=8cm]{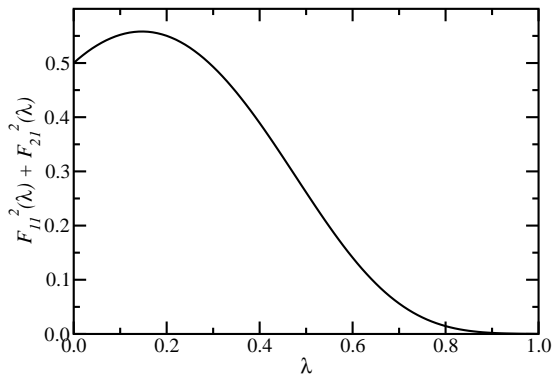}}
\caption{The dimensionless static energy release rate immediately after branching $2\mu G'_s/K_1^2$ as a function of the branching angle $\lambda$, when the primary crack is under mode~I loading.} \label{fig:Gstatic}
\end{figure}

Under in-plane configuration, the exact dependence of the matrix $\mathcal{F}$ on the crack velocity after branching and on the branching angle is not available. However, the exact resolution of the mode~III problem \citep{adda04b} does give an indication about its general behavior, since in many cases, physical aspects of crack propagation of corresponding anti-plane and in-plane configurations are qualitatively similar \citep{Broberg,Freund}. In particular, the results of \citep{adda04b} show that $\mathcal{F}_{33}(\lambda,v')$ depends only weakly on $v'$: the ratio $\mathcal{F}_{33}(\lambda,v')/F_{33}(\lambda)$ is very close to unity (up to $±5\%$) for all values of $\lambda$ and $v'$. We are
then led to assume that this property will also hold for all the matrix elements $\mathcal{F}_{lm}(\lambda,v')$. Therefore, the energy release rate immediately after branching for in-plane configurations is also maximal for branches that propagate quasi-statically ($v'\rightarrow 0$), that is, when $G' = G'_{s}$. Fig.~\ref{fig:Gstatic} shows the dimensionless static energy release rate, $2\mu G'_s/K_1^2\equiv(F_{11}^2+F_{21}^2)$, as a function of the branching angle $\lambda$. Note that this quantity equals $1/2$ for ``zero" branching angle and that it displays a maximum at a nonzero branching angle.

\begin{figure}[ht]
\centerline{\includegraphics[width=8cm]{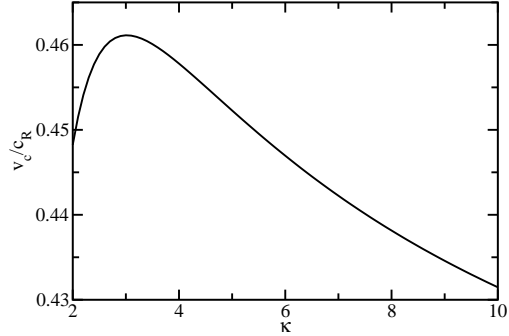}}
\caption{The critical velocity for branching $v_c$ as a function of $\kappa$ for constant fracture energy.}
\label{fig:Vcr1}
\end{figure}

The Griffith's criterion together with the principle of local symmetry (as given by Eq.~(\ref{eq:lambdaCR})) allow to determine direct dynamical properties of the branching instability. To be precise, Eq.~(\ref{eq:Griffith3}) reduces to
\begin{equation}
\frac{\Gamma(0)}{\Gamma(v_c)}g_1(v_c)=F_{11}^2(\lambda_c).
\label{eq:Griffith5}
\end{equation}
This equation selects a critical velocity for branching, denoted by $v_c$. When the fracture energy is velocity independent, Eq.~(\ref{eq:Griffith5}) simplifies to $g_1(v_c)=F_{11}^2(\lambda_c)\simeq0.56$. Obviously, the critical velocity depends on the material properties through $\kappa$ and $c_R$ only, but as shown in Fig.~\ref{fig:Vcr1}, it has a weak dependence on $\kappa$, and can be reasonably taken to be $v_c=0.46 c_R$ - its value for $\kappa \simeq 3$.

\begin{figure}[ht]
\centerline{\includegraphics[width=8cm]{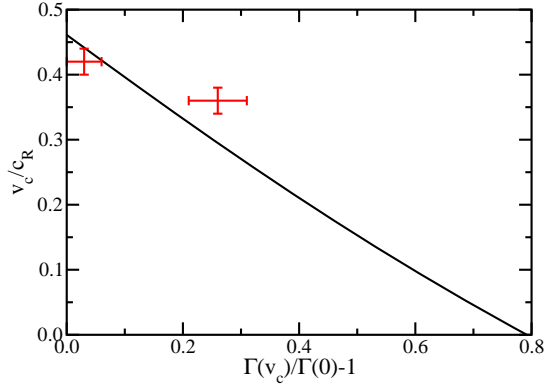}}
\caption{The critical branching velocity for the case of velocity dependent fracture energy and for $\kappa = 3$. The experimental values shown are estimates taken from \citep{Sharon99} for Glass and PMMA.} \label{fig:Vcr2}
\end{figure}

When taking into account the velocity dependence of the fracture energy, Eq.~(\ref{eq:lambdaCR}) which selects the branching angle $\lambda_c$ is not modified, while Eq.~(\ref{eq:Griffith5}) shows that the critical velocity for branching varies. In general, $\Gamma(v)$ is an increasing function of the velocity \citep{Boudet96,Sharon99}. Therefore, the left-hand side of Eq.~(\ref{eq:Griffith5}) decreases faster than in the constant fracture energy case, and the energy balance can thus be achieved at a lower velocity. Although $\Gamma(v)$ can be a nonlinearly dependent function of the crack tip speed, it is only the amount of $\Gamma(v_c)/\Gamma(0)$ which is of importance in determining $v_c$. In Fig.~\ref{fig:Vcr2}, the critical crack tip speed for branching is plotted for different values of $\Gamma(v_c)/\Gamma(0)$ and compared with experimental values for Glass and PMMA as given by \citep{Sharon99}.

It is interesting to mention here that if instead of the principle
of local symmetry, the maximum energy release rate criterion
\citep{Erdogan63} is used, then the equations
determining the critical branching angle $\lambda_c$ and the
critical branching velocity $v_c$ are changed and so do the
resulting values. In that case, $\lambda_c$ is given by the maximum
of $F_{11}^2(\lambda)+F_{21}^2(\lambda)$, and $v_c$ is just the
solution of $g_1(v_c)=F_{11}^2(\lambda_c)+F_{21}^2(\lambda_c)$.
Interestingly, in practice $\lambda_c$ and $v_c$ derived from these
conditions correspond to almost the same values as those obtained from the
principle of local symmetry. This is consistent with other examples
given in the literature, such as the case of kinked cracks treated
in \citep{amle}, where these two criteria yield almost the same
numbers, and in contradiction to \citep{adda05} where somewhat
different results were obtained due to numerical inaccuracies in the
determination of the functions $F_{lm}(\lambda)$.

\subsection{Dynamics of the branches}

The branching picture adopted here, i.e. with a fast moving main crack that stops, branches, and then the branches re-accelerate, might be questionable. Arguments criticizing this scenario rely on the fact that in experiments such full stops are not observed. However, this discrepancy can be related to different effects, such as the three dimensional nature of the experiments or to the fact that real materials are not ideally brittle, and so plasticity could smooth the present picture. Even though velocity fluctuations can be observed, if the crack stops completely then one would expect to see the
branches accelerating gradually from zero velocity to $v_c$ where the next branching event could take place. As we show below, the acceleration of the new branches can be rather large so that the rapid variation of the speed can be easily missed due to lack of temporal experimental resolution.

In order to provide an estimate for the acceleration of the branches' tips, we use Griffith's energy criterion and the principle of local symmetry at each instant of the propagation of the tip of the branch. For a velocity independent fracture energy $\Gamma$, this yields
\begin{eqnarray}
\Gamma &\simeq&  \frac{1}{2\mu} g_1 (v')K'^2_1 (t,v'=0) \nonumber\\
&\simeq&\frac{1}{2\mu }g_1 \left(v'\right)F_{11}^2\left(\lambda_c\right)K_1^2\left[1  + 2
\frac{G_1\left(\lambda_c\right)H_{21}\left(\lambda_c\right) - G_2\left(\lambda_c\right)H_{11}\left(\lambda_c\right)}{H_{21} \left(\lambda_c\right)F_{11}\left(\lambda_c\right)} \frac{T}{K_1}\sqrt{\ell} \right], \label{eq:Griffith6}
\end{eqnarray}
where terms up to order $\sqrt\ell$ in the expansion of the SIFs~(\ref{eq:SIFs}) and the expression for the curvature $a$, Eq. ~(\ref{eq:a}), were used. In addition, knowing that immediately after branching, that is when $\ell\rightarrow0$, the crack speed of the branches is vanishingly small, i.e. $v'\rightarrow0$, Eq.~(\ref{eq:Griffith6}) yields
\begin{equation}
\Gamma=\frac{1}{2\mu} K'^2_1 (0^+,v'=0)=
\frac{1}{2\mu}F_{11}^2(\lambda_c) K_1^2 \;.
\label{eq:Griffith7}
\end{equation}
By equating the last two expressions we get
\begin{equation}
\frac{1-g_1(v')}{g_1(v')} \simeq -0.7 \frac{T}{K_1} \sqrt\ell \;,
\label{eq:Griffith8}
\end{equation}
where the numerical coefficient is computed by using the values of $F_{lm}$, $G_l$ and $H_{lm}$ at $\lambda=\lambda_c$. Eq.~(\ref{eq:Griffith6}) is valid for a small extension of the branch, thus the velocity $v'$ is small and one can safely develop
\begin{equation}
\frac{1-g_1(v')}{g_1(v')} \simeq C(\kappa)\frac{v'}{c_R},
\label{eq:g1_1}
\end{equation}
where $C(\kappa)$ is a numerical coefficient of order unity that can be computed from the asymptotic analysis of $g_1(v)$ as given in \citep{Freund}. For a typical material value $\kappa=3$, one has $C(3)=1.15$. Putting all together, one gets
\begin{equation}
v' = \frac{{d\ell }}{{dt}} \simeq -0.6 \frac{T}{K_1}c_R \sqrt \ell\;,
\label{eq:Griffith9}
\end{equation}
which is a differential equation for $\ell(t)$ that can easily be solved. The quantity $T/K_1$ has been computed in Sec.~\ref{section:T} for the experimental setup of \citep{Fineberg92,Sharon99,Fineberg05}. Using the result $T/K_1=-0.43/\sqrt{W}$ for $\kappa=3$, the length of the branch extension is then given by
\begin{equation}
\ell(t) \simeq 0.017\frac{c_R^2}{W}t^2\;,
\label{eq:Griffith10}
\end{equation}
The branch velocity is then just
\begin{equation}
v'(t) = \frac{{d\ell }}{{dt}} \simeq 0.034\frac{{c_R^2 }}{W}t\;,
\label{eq:Griffith11}
\end{equation}
and finally the acceleration is
\begin{equation}
\frac{{dv' }}{{dt}} \simeq 0.034\frac{c_R^2}{W}\;.
\label{eq:Griffith12}
\end{equation}
It is now obvious that the estimated acceleration of the crack tip
after branching is very large ($c_R\sim 10^3-10^5 m/s$ and $W\sim10^{-2}-10^{-1}m$). This result might explain why even if the crack speed is vanishingly small
immediately after the branching event, it would be difficult to detect it.

\section{Resolution of the static branching problem}

In this section we describe the analytical method we used for the resolution of the elastostatic problem of a long crack with two side branches. First the problem of straight branches is solved, rendering the universal functions $F_{lm}$ and $G_l$. Then, the problem of curved branches is addressed, resulting in $H_{lm}$. The numerical results of this section were already summarized above in Figures~\ref{fig:Flm}, \ref{fig:Gl} and \ref{fig:Hlm}.

\subsection{Straight branches configuration}
\label{section:straight}

\begin{figure}[ht]
\centerline{\includegraphics[width=16cm]{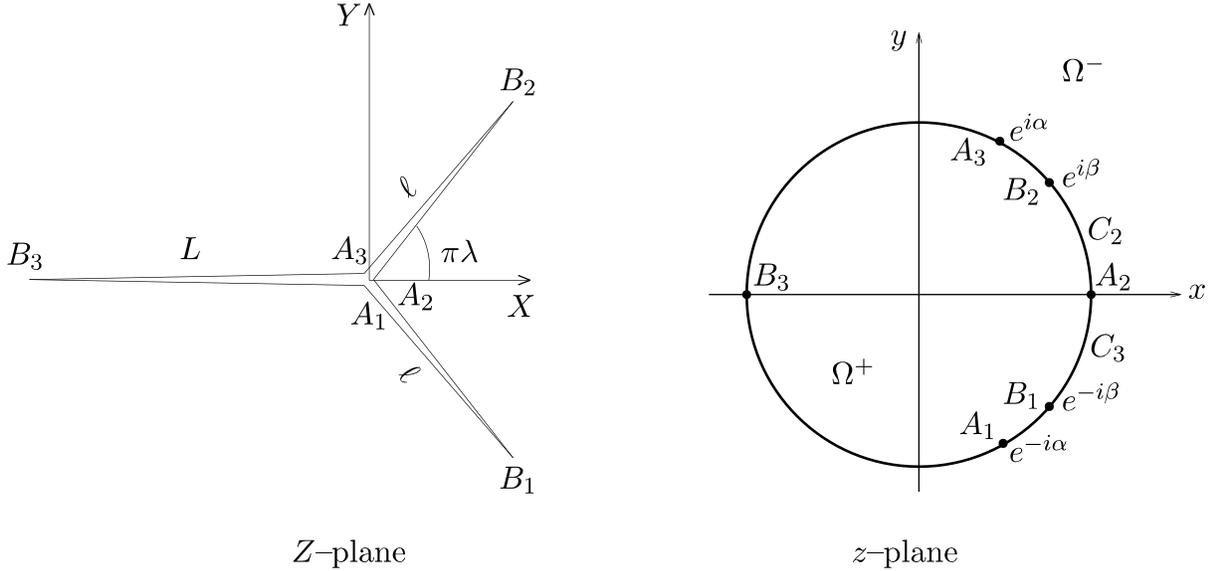}}
\caption{Conformal mapping of a star shaped crack in the $Z$-plane
onto the exterior of the unit circle in the $z$-plane.}
\label{fig:conformal}
\end{figure}

Let us start by giving the general solution of the elastostatic
problem depicted in Fig. \ref{fig:conformal}a. An infinite sheet is
stretched in the presence of a crack contour consisting of a main
crack of length $L$ and two symmetric side branches of equal lengths
$\ell$ emerging from a common origin. The angle between the two
side-branches is denoted by $2\lambda\pi$ with $0<\lambda<1$. In
particular, the case of a main crack with two side-branches of
infinitely small lengths is studied. The elastic potentials of the planar problem
are determined for this geometry and loading. A conformal mapping of the exterior of this star shaped crack into the exterior of a unit circle allows to obtain integral equations for these potentials. Expressions for
the stress intensity factors are derived. In the
following, a detailed resolution of the mixed mode I–-II loading is
presented. Actually, the approach is analogous to the kinked crack
problem which was studied previously in \citep{amle}.

\subsubsection{Conformal Mapping, Potentials}
According to Muskhelishvili \citep{Muskhelishvili}, the stresses and
displacements at a point $Z=X+iY=\omega(z)$ (i.e. a two dimensional
region - see Fig. \ref{fig:conformal}) can be expressed, in the $z$-plane, in terms of the elastic potentials $\Phi(z)$ and $\Psi(z)$. We consider here the case where
traction free boundary conditions are taken on the crack surfaces
and where the loading is given by external stresses
$\sigma_{11}^\infty$ , $\sigma_{22}^\infty$ and
$\sigma_{12}^\infty$ applied at infinity. The goal in this section is to solve for the
Mushkelishvili potentials $\Phi$ and $\Psi$ outside the "star"
shaped crack with two symmetrical straight branches, by using a conformal mapping transformation. The crack is
located in the $Z=X+iY$ space (see Fig. \ref{fig:conformal}), it
corresponds to the curve $\cal C$, and its tips are located at the points
$B_3=-L$ (tip of the left end of the original straight crack), $B_2=\ell
\exp (i\lambda\pi)$ and $B_1=\ell \exp(-i\lambda\pi)$ (tips of the
branches). The region exterior to the crack is named $\Omega^-$. The
potentials satisfy the following equation on the crack line $\cal C$:
\begin{equation}
\Phi(Z)+Z\overline{\Phi'(Z)}+\overline{\Psi(Z)}=Const \; ,
\label{PP}
\end{equation}
and the following boundary conditions at infinity:
\begin{eqnarray}
\Phi(Z)=\Gamma Z, \\
\Psi(Z)=\Gamma' Z
\end{eqnarray}
with $\Gamma \equiv (\sigma_{11}^{\infty}+\sigma_{22}^{\infty})/4$,
$\Gamma' \equiv
(\sigma_{22}^{\infty}-\sigma_{11}^{\infty})/2+i\sigma_{12}^{\infty}$,
i.e. given in terms of the stresses at infinity.

The following conformal mapping \citep{Smith68}:
\begin{equation}
Z=\omega(z)=
\frac{C}{z}\left(z-z_1\right)\left(z-z_3\right)\left(\frac{z-z_2}{z-z_1}\right)^\lambda
\left(\frac{z-z_2}{z-z_3}\right)^\lambda
\end{equation}
maps the exterior of the "star" shaped crack described by the points
$A_1, B_1, A_2,B_2, A_3, B_3$ in the $Z$-plane (see FIG.
\ref{fig:conformal}a) to the exterior of a unit circle in the
$z$-plane, with corresponding points $z_1, y_1, z_2, y_2, z_3, y_3$
(see FIG. \ref{fig:conformal}b). These special points are located
at: $z_1=e^{-i\alpha}$, $z_2=1$, $z_3=e^{i\alpha}$,
$y_1=e^{-i\beta}$, $y_2=e^{i\beta}$ and $y_3=-1$. The constants
$C$, $\alpha$ and $\beta$ are given in terms of the lengths $L$ of the
main crack, and $\ell$ of each branch through the following equations:
\bqe
C & = & \frac{L}{4}\left[\cos \left(\alpha/2\right)\right]^{2(\lambda-1)}
\label{Clbb} \\
\ell & = &  \frac{L}{2} \lambda \left(1-\cos \alpha\right) \left\{ \frac{2\left(1-\lambda\right)}{\lambda(1+\cos \alpha)} \right\}^{1-\lambda} \\
\sin(\beta/2) & = & \sqrt{\lambda} \sin (\alpha/2)
\label{Clbe}
\eqe
 In the
$z$-plane, Eq. (\ref{PP}) for the elastic potentials reads:
\begin{equation}
\phi(z)+\frac{\omega(z)}{\overline{\omega'(z)}}
\overline{\phi'(z)}+\overline{\psi(z)}=Const \; , \label{pp}
\end{equation}
($\phi(z)=\Phi(Z)$, $\psi(z)=\Psi(Z)$)
and the boundary conditions at infinity become $\phi(z)=\Gamma Cz$,
$\psi(z)=\Gamma' C z$. The quantity
$\omega(z)/\overline{\omega'(z)}$ appearing in Eq. (\ref{pp}) takes
different values in different sections of the crack line:
\begin{equation}
\frac{\omega(z)}{\overline{\omega'(z)}}=\left\{-1+(1-e^{-i2\lambda\pi})I_1(z)+(1-e^{i
2\lambda\pi})I_2(z)\right\}Q(z) \; ,
\end{equation}
with $I_{1,2}(z)=1$ if $z$ belongs to ${\cal C}_{1,2}$ respectively, and
zero otherwise. ${\cal C}_{1,2}$ are the branches of the crack, i.e. these
curves unite the points $A_1, A_2$ and $A_2,A_3$ respectively (see
Fig. \ref{fig:conformal}b); and the function $Q(z)$ is the
following:
\begin{equation}
Q(z) =
\frac{(z-e^{-i\alpha})(z-e^{i\alpha})(z-1)}{z(z+1)(z-e^{i\beta})(z-e^{-i\beta})}
\label{Qz}.
\end{equation}
In order to solve equation (\ref{pp}), the following Lemma will be
used: if $f$ and $g$ are complex functions defined and continuous in
$\Omega^- \cup \cal C$, analytic on $\Omega^-$, (including the point
at infinity) and such that $f(z)=\overline{g(z)}$ for $z \in \cal C$,
then $f$ and $g$ are constants and conjugate to each other. Thus,
Eq. (\ref{pp}) will be written in the previous form, i.e. as an
equality between an analytic function and the complex conjugate of
another analytic function. This is accomplished by defining:
\begin{equation}
\chi_{1,2}(z) \equiv \frac{(1-e^{\mp i 2\lambda\pi})}{2\pi i}
\int_{{\cal C}_{1,2}} d\eta \frac{Q(\eta)\overline{\phi'(\eta)}}{(\eta-z)} \label{c23}
\end{equation}
If $z^+$
and $z^-$ represent points just inside and outside of the unit
circle respectively, then by Plemelj's formula:
\begin{equation}
\chi_{1,2}(z^+)-\chi_{1,2}(z^-)=(1-e^{\mp i
2\lambda\pi})Q(z)\overline{\phi'(z)}
\label{chi12}
\end{equation}
if $z^+, z^-$ are on ${\cal C}_1$ or ${\cal C}_2$ respectively, and zero otherwise.
From equations (\ref{Qz}),
(\ref{c23}) one sees that convergence issues at the points
$e^{i\beta}$ and $e^{-i\beta}$ (poles of $Q(z)$) can be addressed by
understanding these integrals with these poles slightly displaced
into $\Omega^-$ (i.e. $\beta \rightarrow \beta \mp i \epsilon$ in
the poles at $e^{\pm i \beta}$, respectively). Using these definitions and results
in Eq. (\ref{pp}), the latter becomes:
\begin{equation}
\phi(z)-\chi_1(z^-)-\chi_2(z^-)=\overline{
Q_*(z)\phi'(z)-\chi_{1*}(z^-)-\chi_{2*}(z^-)-\psi(z)}+Const
\label{nc}
\end{equation}
where $f_*(z) \equiv \overline{f(1/\overline{z})}$ is an analytic
function of $z$ if $f(z)$ is analytic. Thus,
\begin{equation}
Q_*(z) =
-\frac{z(z-e^{i\alpha})(z-e^{-i\alpha})(z-1)}{(z-e^{i\beta})(z-e^{-i\beta})(z+1)}
\label{qast}
\end{equation}
Notice that if $Q(z)$ has poles at $z=e^{\epsilon}e^{\pm i \beta}$
(i.e., in $\Omega^-$), $Q_*(z)$ has corresponding poles at
$z=e^{-\epsilon}e^{\pm i\beta}$ (i.e., in $\Omega^+$). The condition
of analyticity at $\infty$ of the Lemma has to be examined for Eq.
(\ref{nc}). Indeed the left hand side and the conjugate of the right
hand side of the latter equation behave at infinity as:
\begin{eqnarray}
\phi(z)-\chi_1(z)-\chi_2(z) & \simeq & \Gamma C z  \\
Q_*(z)\phi'(z)-\chi_{1*}(z)-\chi_{2*}(z)-\psi(z) & \simeq &
-(\Gamma+\Gamma')Cz.
\end{eqnarray}
The behaviors at infinity are regularized (linear terms in $z$ are
eliminated there) if one adds the terms $-\Gamma C
z+(\Gamma+\overline{\Gamma'})\overline{C}/z$ and $\overline{-\Gamma
\overline{C}/z+(\Gamma+\Gamma')Cz}$ to the left hand side and
right hand side of Eq. (\ref{nc}) respectively, which then becomes:
\begin{eqnarray}
\phi(z) &-& \chi_1(z^-)-\chi_2(z^-) - \Gamma C z
 +(\Gamma+\overline{\Gamma'})\overline{C}/z \nonumber \\
 & = & \overline{Q_*(z)\phi'(z)-\chi_{1*}(z^-)-\chi_{2*}(z^-)-\psi(z)
-\Gamma \overline{C}/z+(\Gamma+\Gamma')Cz}+Const \label{eqle}
\end{eqnarray}

Applying the Lemma to Eq. (\ref{eqle}) (left hand side), and
using Eq. (\ref{chi12}), one gets:
\begin{eqnarray}
\phi (z) & = &  \Gamma C z-\frac{(\Gamma+\overline{\Gamma'})\overline{C}}{z} \nonumber \\
& & + \frac{(1-e^{-2i\lambda\pi})}{2\pi i} \int_{{\cal C}_1} d\eta
\frac{Q(\eta)\overline{\phi'(\eta)}}{(\eta-z)}
+\frac{(1-e^{2i\lambda\pi})}{2\pi i} \int_{{\cal C}_2} d\eta
\frac{Q(\eta)\overline{\phi'(\eta)}}{(\eta-z)} + Const \; ,
\end{eqnarray}
i.e. an integral equation for $\phi'(z)$ which can be written in the form:
\begin{equation}
\phi'(z)=\phi_0'(z)+\mathcal{L}(\phi'(z)) \; , \label{fpo}
\end{equation}
with
\begin{equation}
\phi_0'(z) \equiv  \Gamma C
+\frac{(\Gamma+\overline{\Gamma'})\overline{C}}{z^2} \; ,
\end{equation}
and the operator $\mathcal{L}$ is defined through:
\begin{eqnarray}
\mathcal{L}(f(z)) & = & \frac{(1-e^{-2i\lambda\pi})}{2\pi i}
\int_{{\cal C}_1} d\eta \frac{
(\eta-e^{-i\alpha})(\eta-e^{i\alpha})(\eta-1)\overline{f(\eta)}}{
\eta(\eta+1)(\eta-e^{i\beta})(\eta-e^{-i\beta})(\eta-z)^2} \nonumber \\
&+& \frac{(1-e^{2i\lambda\pi})}{2\pi i} \int_{{\cal C}_2} d\eta
\frac{(\eta-e^{-i\alpha})(\eta-e^{i\alpha})(\eta-1)\overline{f(\eta)}}{
\eta(\eta+1)(\eta-e^{i\beta})(\eta-e^{-i\beta})(\eta-z)^2} + Const
\label{Lop}
\end{eqnarray}

\subsubsection{Expansion in powers of the crack extension length}

In this section we specialize to the case $\ell \rightarrow 0$,
i.e. a situation with a long macroscopic crack with two micro-cracks right after branching. First, we write in the limit $\ell \rightarrow 0$ an asymptotic
expression for the constants $C$, $\alpha$ and $\beta$ that follow
from Eqs. (\ref{Clbb})-(\ref{Clbe}):
\begin{eqnarray}
C & = & \frac{L}{4}+O(\ell) \\
\alpha & = & \sqrt{\frac{4}{(1-\lambda)L}}\left(\frac{1-\lambda}{\lambda}\right)^{\lambda/2}\sqrt{\ell}  \label{alfa} \\
\beta & = & \sqrt{\lambda} \alpha \label{asym}
\end{eqnarray}
Writing
\begin{equation}
z=e^{i \alpha \zeta} \qquad \mbox{and} \qquad \phi'(z) = e^{-i\alpha
\zeta}\left[\sqrt{L}U(\zeta) +\alpha L
V(\zeta)+O(\alpha^2)\right],
\label{eq:phi'}
\end{equation}
equation (\ref{fpo}) becomes to order $\alpha$ (Notice that
$\Omega^-$ corresponds to the lower half plane in the complex
$\zeta$ plane, i.e. the poles at $z=e^{\pm i \beta}(1 \mp i
\epsilon)$ are now located at $\zeta=\mp \sqrt{\lambda}-i
\tilde{\epsilon}$):
\begin{eqnarray}
\frac{U(\zeta)}{\sqrt{L}}+\alpha V(\zeta) & = & \frac12
\left(\Gamma+\frac{\overline{\Gamma'}}{2}\right) -i \alpha
\frac{\overline{\Gamma'}}{4}\zeta +
\frac{(1-e^{-i2\lambda\pi})}{4\pi i} \int_{-1}^0 dh
\frac{h (h^2-1) \left(\overline{U(h)}/\sqrt{L}+\alpha
\overline{V(h)}\right)}{(h^2-\lambda)(h-\zeta)^2} \nonumber \\
& + & \frac{(1-e^{i2\lambda\pi})}{4\pi i}
\int_0^1 dh \frac{h (h^2-1)
\left(\overline{U(h)}/\sqrt{L}+\alpha
\overline{V(h)}\right)}{(h^2-\lambda)(h-\zeta)^2}
\end{eqnarray}
which yields the following equations for $U(\zeta)$ and
$V(\zeta)$:
\begin{eqnarray}
U(\zeta) & = & U^{(0)}(\zeta)+\mathcal{A}U(\zeta) \label{eq:U}\\
V(\zeta) & = & V^{(0)}(\zeta)+\mathcal{A}V(\zeta) \label{eq:V}
\end{eqnarray}
where the functions $U^{(0)}$ and $V^{(0)}$ and the operator
$\mathcal{A}$ are the following:
\begin{eqnarray}
U^{(0)}(\zeta) & \equiv &
\frac{\sqrt{L}}{2}
\left(\Gamma+\frac{\overline{\Gamma'}}{2}\right)=\frac{\left(K_1-iK_2\right)}
{\sqrt{8 \pi}} \label{eq:U0}\\
V^{(0)}(\zeta) & \equiv & -i \frac{\overline{\Gamma'}}{4} \zeta =
\left(-\sigma_{12}^{\infty}+i\frac{T}{2}\right)\frac{\zeta}{4} \;, \label{eq:V0}\\
\mathcal{A}f(\zeta) & \equiv & \frac{(1-e^{-i2\lambda\pi})}{4\pi i} \int_{-1}^0 dh
\frac{h (h^2-1)\overline{f(h)}}{(h^2-\lambda)(h-\zeta)^2} + \frac{(1-e^{i2\lambda\pi})}{4\pi i}
\int_0^1 dh \frac{h (h^2-1) \overline{f(h)}}{(h^2-\lambda)(h-\zeta)^2}
\label{Afm}
\end{eqnarray}
where
\begin{equation}
K_1-iK_2=\left(\sigma_{22}^{\infty}-i\sigma_{12}^{\infty}\right)\sqrt{\frac{\pi
L}{2}} \label{k0}
\end{equation}
are the stress intensity factors of the original single crack of
length $L$ under the same loading, and $T$ is the non-singular
stress ($\sigma_{xx}$) at the original crack tip.

Andersson's formula \citep{Andersson69} for the stress intensity factors can be applied
at the upper crack tip (one should get an analogous result at the
lower tip), as follows:
\begin{equation}
K'_1(\ell)-i K'_2(\ell)
=2\sqrt{\pi}
\phi'(e^{i\beta})e^{-i\delta/2}/\sqrt{\omega''(e^{i\beta})}
\label{And}
\end{equation}
where $\delta$ is the angle between the $X$ axis and the tangent
to the crack at its tip, i.e. $\pi \lambda$ in this case. To first
order in $\alpha$, we get
\begin{equation}
\omega''(e^{i\beta}) \simeq L\left(1-2i \sqrt{\lambda}
\alpha\right)\left(\frac{\lambda}{1-\lambda}\right)^\lambda
e^{i\lambda\pi}
\end{equation}
Also, writing the stress intensity factors up to
order $\sqrt{\ell}$ as:
\begin{equation}
K'_1(\ell)-i K'_2(\ell) =K_1^*-i K_2^*
+\left(K_1^{(1/2)}-i K_2^{(1/2)}\right)\sqrt{\ell} \; ,
\end{equation}
one obtains:
\begin{eqnarray}
K_1^*-i K_2^* & = & 2\sqrt{\pi}e^{-i\lambda \pi}\left(\frac{1-\lambda}{\lambda}\right)^{\lambda/2}U\left(\sqrt{\lambda}\right) \label{Flm1}\\
K_1^{(1/2)}-i K_2^{(1/2)} & = & 4 \sqrt{\frac{\pi}{1-\lambda}}
e^{-i\lambda \pi}
\left(\frac{1-\lambda}{\lambda}\right)^{\lambda}V\left(\sqrt{\lambda}\right)
\label{Gl1}
\end{eqnarray}

Equations (\ref{eq:U})-(\ref{k0}), (\ref{Flm1}) and (\ref{Gl1}) show
that the $K_l^*$'s and $K_l^{(1/2)}$'s can be determined
independently, i.e. the $K_l^{(1/2)}$'s depend only on the function
$U$ which can be found from Eqs.~(\ref{eq:U}),(\ref{eq:U0}) and
(\ref{Afm}) where the function $V$ does not appear. Similarly, the
$K_l^{(1/2)}$'s can be found from Eqs.~(\ref{eq:V}),(\ref{eq:V0})
and (\ref{Afm}) where the function $U$ does not appear. This
remarkable property holds only in the limit $\ell \rightarrow 0$,
and is easily evidenced thanks to the addition of the term
$e^{-i\alpha\zeta}$ in the expansion for $\phi'(z)$ (Eq.~(\ref{eq:phi'})).
Also, since the function $U^{(0)}$ depends on the three components
of the stress tensor at infinity only through two parameters, namely
the SIF's at the initial crack tip, the same holds for $U$ and for
the $K_l^*$s, i.e. they depend on $K_1, K_2$ (this property is again true only in the limit $\ell
\rightarrow 0$). In the same spirit, one could have thought that
since $V^{(0)}$ depends on $T$ and on $\sigma_{12}^{\infty}$ the
same is true for $V$, namely that it depends on both quantities $T$
and $\sigma_{12}^{\infty}$. However, this fact contradicts a
universality prediction presented in \citep{amle} which states
that the $K_l^{(1/2)}$'s should only depend on the $T$-stress. In
fact, the contradiction is only apparent since it can be shown that
the part of $V$ which arises from $\sigma_{12}^{\infty}$ (i.e. a $V$
that solves Eqs.~(\ref{eq:V}), (\ref{eq:V0}) with $T=0$) has the following closed form
\begin{equation}
\left[V\left(
\zeta \right)\right]_{T=0} =  - \frac{\sigma _{12}^\infty  }
{4}\frac{(\zeta ^2 - \lambda) } {\zeta }\left( {\frac{\zeta ^2 }
{\zeta ^2  - 1}} \right)^\lambda  \qquad (\zeta \in
\Pi^-)\label{exactV}.
\end{equation}
Now, it is easily seen that the function $\left[V\left( \zeta
\right)\right]_{T=0}$ is zero at the point $\zeta=\sqrt{\lambda}$,
so that $\sigma _{12}^\infty$ does not contribute to the SIFs
$K_l^{(1/2)}$'s which are given by expression (\ref{Gl1}). This
observation shows that the universality predicted by \citep{amle} is respected. The exact result of Eq.~(\ref{exactV}) serves as a useful
check on the correctness of any numerical analysis of these
equations.

\subsubsection{Numerical considerations}
\begin{figure}
\centerline{\includegraphics[width=8cm]{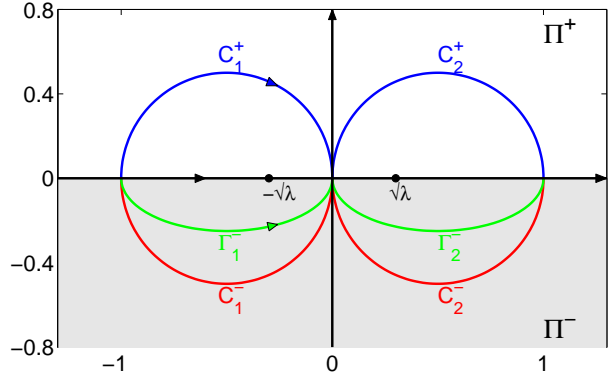}}
\caption{The $\zeta$-plane with some useful contours.}
\label{contours}
\end{figure}

Unfortunately, an analytical solution such as (\ref{exactV}) was not
found for the function $U$ and for that part of $V$ which is
proportional to $T$. Therefore, in order to determine the functions
$F_{lm}(\lambda)$ and $G_l(\lambda)$ it is necessary to solve
numerically for $U$ and $V$. Let us begin with $U$, where a useful
decomposition is given by
\begin{equation}
U(\zeta) = \frac{1}{\sqrt{8\pi}} (K_1 U_1( \zeta)
 - i K_2 U_2 ( \zeta)). \label{decomp:U}
\end{equation}
Eq. (\ref{eq:U}) is now decomposed
into two equations
\begin{eqnarray}
U_{1,2} \left( \zeta  \right) = 1 & \pm & \frac{{1 - e^{ - 2i\lambda \pi } }} {{4i\pi }}\int_{C_1^+}
{\frac{{\eta \left( {\eta ^2  - 1} \right)}} {(\eta ^2 - \lambda )}\frac{{\overline{U_{1,2}} \left( \eta \right)d\eta
}} {{\left( {\eta  - \zeta } \right)^2 }}} \nonumber\\
&\pm& \frac{{1 - e^{2i\lambda \pi } }} {{4i\pi }}\int_{C_2^+}
{\frac{{\eta \left( {\eta ^2 - 1} \right)}} {(\eta ^2
- \lambda )}\frac{{\overline{U_{1,2}} \left( \eta \right)d\eta}} {{\left( {\eta - \zeta } \right)^2 }}}
\label{eq:U12},
\end{eqnarray}
where we deformed the integration paths away from the poles
$\pm \sqrt{\lambda} -i\epsilon$ onto two semi-circles denoted
$C_1^+$ and $C_2^+$ respectively (i.e. $|\zeta
\pm \frac{1}{2}|=\frac{1}{2}$, Im $\zeta > 0$, oriented from $-1$
through $0$, and from $0$ to $1$ (see Fig. \ref{contours})).

Notice that the function $\overline{f}(z) \equiv \overline{f(\overline{z})}$
has been introduced, which is analytic when $f(z)$ is analytic ($\overline{f}(z)$ coincides with $\overline{f(z)}$ on the real axis, and it is its analytic continuation into the rest of the complex plane).
Eqs. (\ref{eq:U12}) can be solved numerically using an iterative
method. Beginning with the non-homogenous term in the equations
$U_{1,2}^{(0)}=1$, we can iterate using the operator $\mathcal{A}$
(given in eq. (\ref{Afm})) and obtain the following formal
solutions
\begin{eqnarray}
U_1 \left( \zeta \right) &=& \sum\limits_{n =
0}^\infty {\mathcal{A}^n U_1^{(0)} \left( \zeta  \right)} \label{eq:U1sol}\\
U_2 \left( \zeta  \right) &=& \sum\limits_{n = 0}^\infty  {\left(
{ - 1} \right)^n \mathcal{A}^n U_2^{(0)} \left( \zeta \right)}
\label{eq:U2sol}
\end{eqnarray}
which can be seen to converge, and to yield the required result. The
convergence is due to the fact that the operator $\mathcal{A}$ is
contracting in the space of functions defined and continuous on
$C_{1,2}^-$ with, at most, weak singularities at $0, \pm 1$ (see
Appendix A for a proof).

The simplest way is to compute the functions $U_1(\zeta)$ and
$U_2(\zeta)$ on the lower semi-circles $C_1^-$ and $C_2^-$
respectively (see Fig. \ref{contours}) since then
$\overline{U_1}(\zeta)$ and $\overline{U_2}(\zeta)$ which are needed
in the integrals (\ref{eq:U12}) are trivially obtained on $C_1^+
\cup C_2^+$ by conjugation of $U_1(\overline{\zeta})$ and $U_2(\overline{\zeta})$. In
practice, we parametrized each of these two functions by two
functions defined on $[0,\pi]$, namely $U_{1,2}^a(\gamma) \equiv
U_{1,2}(\frac{-1-e^{i\gamma}}{2})$ on $C_1^-$ and $U_{1,2}^b(\gamma)
\equiv U_{1,2}(\frac{1-e^{i\gamma}}{2})$ on $C_2^-$ (with $\gamma
\in [0,\pi]$). In order to be more specific we write down explicitly
the four singular integral equations for $U_{1,2}^{a,b}$ that we
solved numerically
\begin{eqnarray}
 U_{1,2}^a \left( \gamma  \right) = 1 & \pm & \frac{{1 - e^{ - 2i\lambda \pi } }}{{8\pi }}\int_0^\pi  {\frac{{\left( { - {\textstyle{{1 + e^{ - i\theta } } \over 2}}} \right)\left[ {\left( { - {\textstyle{{1 + e^{ - i\theta } } \over 2}}} \right)^2  - 1} \right]}}{{\left( { - {\textstyle{{1 + e^{ - i\theta } } \over 2}}} \right)^2  - \lambda }}\frac{{\overline {U_{1,2}^a \left( \theta  \right)} e^{ - i\theta } d\theta }}{{\left[ {\left( { - {\textstyle{{1 + e^{ - i\theta } } \over 2}}} \right) - \zeta } \right]^2 }}}  \nonumber \\
& \pm & \frac{{1 - e^{2i\lambda \pi } }}{{8\pi }}\int_0^\pi  {\frac{{\left( {{\textstyle{{1 - e^{ - i\theta } } \over 2}}} \right)\left[ {\left( {{\textstyle{{1 - e^{ - i\theta } } \over 2}}} \right)^2  - 1} \right]}}{{\left( {{\textstyle{{1 - e^{ - i\theta } } \over 2}}} \right)^2  - \lambda }}\frac{{\overline {U_{1,2}^b \left( \theta  \right)} e^{ - i\theta } d\theta }}{{\left[ {\left( {{\textstyle{{1 - e^{ - i\theta } } \over 2}}} \right) - \zeta } \right]^2 }}}  \label{eq:U12a} \\
 U_{1,2}^b \left( \gamma  \right) = 1 & \pm & \frac{{1 - e^{ - 2i\lambda \pi } }}{{8\pi }}\int_0^\pi  {\frac{{\left( { - {\textstyle{{1 + e^{ - i\theta } } \over 2}}} \right)\left[ {\left( { - {\textstyle{{1 + e^{ - i\theta } } \over 2}}} \right)^2  - 1} \right]}}{{\left( { - {\textstyle{{1 + e^{ - i\theta } } \over 2}}} \right)^2  - \lambda }}\frac{{\overline {U_{1,2}^a \left( \theta  \right)} e^{ - i\theta } d\theta }}{{\left[ { - {\textstyle{{1 + e^{ - i\theta } } \over 2}} - {\textstyle{{1 - e^{i\gamma } } \over 2}}} \right]^2 }}}  \nonumber\\
& \pm & \frac{{1 - e^{2i\lambda \pi } }}{{8\pi }}\int_0^\pi  {\frac{{\left( {{\textstyle{{1 - e^{ - i\theta } } \over 2}}} \right)\left[ {\left( {{\textstyle{{1 - e^{ - i\theta } } \over 2}}} \right)^2  - 1} \right]}}{{\left( {{\textstyle{{1 - e^{ - i\theta } } \over 2}}} \right)^2  - \lambda }}\frac{{\overline {U_{1,2}^b \left( \theta  \right)} e^{ - i\theta } d\theta }}{{\left[ {{\textstyle{{1 - e^{ - i\theta } } \over 2}} - {\textstyle{{1 - e^{i\gamma } } \over 2}}} \right]^2 }}} \label{eq:U12b}.
\end{eqnarray}

Once $U_1(\zeta)$ and $U_2(\zeta)$ are known on the lower
semi-circles, we use eqs. (\ref{eq:U12}) once again to obtain
$U_1(\sqrt{\lambda})$ and $U_2(\sqrt{\lambda})$. Then, using
definition (\ref{eq:SIFs}) together with eqs. (\ref{Flm1}) and
(\ref{decomp:U}) we can extract the $F_{lm}$'s, namely the matrix
elements relating the stress intensity factors immediately after
branching at the tip $B_2$ of the infinitely small side-branch to
the stress intensity factors of the main crack of length $L$ in the
absence of the side-branches at the leading order
\begin{eqnarray}
 F_{11} \left( \lambda  \right) &=& \frac{1}{{\sqrt 2 }}\left( {\frac{{1 - \lambda }}{\lambda }} \right)^{{\lambda  \mathord{\left/
 {\vphantom {\lambda  2}} \right.
 \kern-\nulldelimiterspace} 2}} \Re \left[ {e^{ - i\lambda \pi } U_1 \left( {\sqrt \lambda  } \right)} \right] \\
 F_{21} \left( \lambda  \right) &=&  - \frac{1}{{\sqrt 2 }}\left( {\frac{{1 - \lambda }}{\lambda }} \right)^{{\lambda  \mathord{\left/
 {\vphantom {\lambda  2}} \right.
 \kern-\nulldelimiterspace} 2}} \Im \left[ {e^{ - i\lambda \pi } U_1 \left( {\sqrt \lambda  } \right)} \right] \\
 F_{12} \left( \lambda  \right) &=& \frac{1}{{\sqrt 2 }}\left( {\frac{{1 - \lambda }}{\lambda }} \right)^{{\lambda  \mathord{\left/
 {\vphantom {\lambda  2}} \right.
 \kern-\nulldelimiterspace} 2}} \Im \left[ {e^{ - i\lambda \pi } U_2 \left( {\sqrt \lambda  } \right)} \right] \\
 F_{22} \left( \lambda  \right) &=& \frac{1}{{\sqrt 2 }}\left( {\frac{{1 - \lambda }}{\lambda }} \right)^{{\lambda  \mathord{\left/
 {\vphantom {\lambda  2}} \right.
 \kern-\nulldelimiterspace} 2}} \Re \left[ {e^{ - i\lambda \pi } U_2 \left( {\sqrt \lambda  } \right)} \right]
\label{Flm2}.
\end{eqnarray}
Results following from a numerical calculation of these functions
were presented above in Fig. \ref{fig:Flm}. In that figure, we
superimposed the results obtained in \citep{amle} for the case of a
kinked crack (with the same angle) for the sake of comparison.

The next stage is to get the $G_l$'s. For that we need to solve eq.
(\ref{eq:V}). Recall that we can dismiss the term
$\sigma_{12}^\infty$ since it will not contribute to the $G_l$'s
(due to the exact result we presented in eq. (\ref{exactV})), and so
by denoting
\begin{equation}
V\left( \zeta  \right) = \frac{T}{8}\hat V\left( \zeta \right),
\label{decomp:V}
\end{equation}
we then need to solve the following equation
\begin{equation}
\hat V\left( \zeta  \right) = i\zeta  + \frac{{1 - e^{ - 2i\lambda
\pi } }}{{4i\pi }}\int_{C_1^+} {\frac{{\eta \left(
{\eta ^2  - 1} \right)}}{{(\eta ^2  - \lambda)
}}\frac{{\overline{\hat V}\left( \eta  \right)d\eta
}}{{\left( {\eta  - \zeta } \right)^2 }}}  + \frac{{1 -
e^{2i\lambda \pi } }}{{4i\pi }}\int_{C_2^+} {\frac{{\eta
\left( {\eta ^2  - 1} \right)}}{{(\eta ^2  - \lambda)
}}\frac{{\overline{\hat V}\left( \eta  \right)d\eta
}}{{\left( {\eta  - \zeta } \right)^2 }}} \label{eq:Vhat}.
\end{equation}
This equation can be solved using the same iterative method as
above, by taking the nonhomogenous term in the equation $\hat
V^{(0)} (\zeta)= i\zeta$ and iterating it using the operator
$\mathcal{A}$
\begin{equation}
\hat V\left( \zeta  \right) = \sum\limits_{n = 0}^\infty
{\mathcal{A}^n \hat V^{\left( 0 \right)} \left( \zeta  \right)}
\label{eq:Vsol}.
\end{equation}
In order to solve this we apply exactly the same procedure as for
$U_{1,2}(\zeta)$, namely deforming the integration contours to
$C_1^+ \cup C_2^+$, and solving for $\hat V(\zeta)$ along the
lower semi-circles $\zeta \in C_1^- \cup C_2^-$. As before, we
parameterize $\hat V(\zeta)$ using two functions defined on
$[0,\pi]$, namely $\hat V^a(\gamma) \equiv \hat
V(\frac{-1-e^{i\gamma}}{2})$ on $C_1^-$ and $\hat V^b(\gamma) \equiv
\hat V(\frac{1-e^{i\gamma}}{2})$ on $C_2^-$ (with $\gamma \in
[0,\pi]$). This results in two singular integral equations for $\hat
V^{a,b}$ that can be solved, and finally yield the required $\hat V
(\sqrt{\lambda})$. Then, using definition (\ref{eq:SIFs}) and Eqs.~(\ref{Gl1}) and (\ref{decomp:V}) we can extract the following
expressions for the $G_l$'s
\begin{eqnarray}
 G_1  &=& \frac{1}{2}\sqrt {\frac{\pi }{{1 - \lambda }}} \left( {\frac{{1 - \lambda }}{\lambda }} \right)^\lambda  \Re \left\{ {e^{ - i\lambda \pi } \hat V\left( {\sqrt \lambda  } \right)} \right\} \\
 G_2  &=&  - \frac{1}{2}\sqrt {\frac{\pi }{{1 - \lambda }}} \left( {\frac{{1 - \lambda }}{\lambda }} \right)^\lambda  \Im \left\{ {e^{ - i\lambda \pi } \hat V\left( {\sqrt \lambda  } \right)} \right\}
\label{Gl2}.
\end{eqnarray}
Results emanating from a numerical resolution of $\hat V$
were presented above in Fig.~\ref{fig:Gl}. In that figure, we
superimposed the results obtained in \citep{amle} for the case of a
kinked crack (with the same angle) for the sake of comparison.

\subsection{Curved branched extensions}
\label{section:curved}

\begin{figure}[ht]
\centerline{\includegraphics[width=8cm]{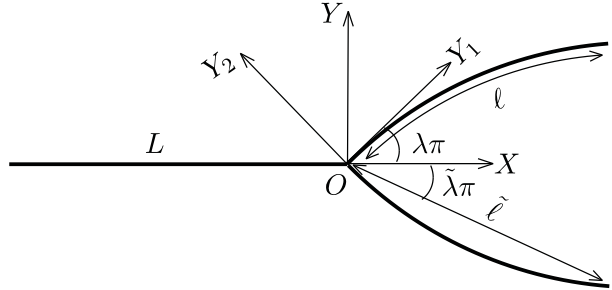}}
\caption{Schematic representation of a straight crack with
symmetrically branched curved extensions. The fictitious straight
crack around which the perturbation expansion is performed is drawn
on the lower branch.} \label{fig:curved}
\end{figure}

We now consider the problem of the curved extensions defined in Fig.~\ref{fig:curved} with the aim to determine the $H_{lm}$'s (defined
by Eq.~(\ref{eq:SIFs})). The curved extension of one branch is
described in terms of the coordinates $Y_{1,2}$, where $Y_1$
is the axis parallel to the initial slope of the branch, that forms an
angle $\pi \lambda$ with the $X$ axis ($Y_2$ is
perpendicular to $Y_1$). In this frame of reference the shape of the
extension can be written as \citep{Leblond89,amle}
\begin{equation}
Y_2=a Y_1^{3/2}+O\left(Y_1^2\right),
\end{equation}
where $a$ is a curvature parameter. A fictitious branch is defined as a straight line joining the beginning and the end of the branches, it forms an angle $\pi \tilde\lambda$ with the $X$ axis.
The length of the fictitious
branch is denoted $\tilde{\ell}$, and $\ell$ is the length of the
curved extension. Expansions of $\tilde{\ell}$ and $\tilde{\lambda}$
are:
\begin{eqnarray}
\tilde{\ell} & = & \ell + O\left(\ell^2\right) \label{eq:lTilde}\\
\tilde{\lambda} & = & \lambda+\frac{a}{\pi}\sqrt{\ell}.
\label{eq:lambdaTilde}
\end{eqnarray}

We will solve below for the Muskhelishvili potentials by a first
order perturbative procedure, with the curvature $a$ being a small parameter. This analysis will yield an expression for the SIFs of the curved branches as functions of the fictitious
variables
$\tilde{\ell}$ and $\tilde{\lambda}$. This expression will not yield directly
the required $H_{lm}$'s defined in Eq.~(\ref{eq:SIFs}), since
one needs to express the latter SIF's in terms of an expansion in $\sqrt{\ell}$ and as
functions of $\lambda$ by simply changing variables from $\tilde{\ell}$, $\tilde{\lambda}$
to $\ell$, $\lambda$. It is therefore necessary to rewrite Eq.~(\ref{eq:SIFs}) in terms of $\tilde{\ell}$ and $\tilde{\lambda}$
using Eqs.~(\ref{eq:lTilde})-(\ref{eq:lambdaTilde}). Following \citep{amle} we get
\begin{eqnarray}
K'_l(\ell) &=& \sum_{m=1,2}F_{lm}(\lambda) K_m + \sum_{m=1,2}
\left[G_m(\lambda) T \delta_{lm} + a H_{lm}(\lambda) K_m \right]
\sqrt{\ell} + O(\ell) \nonumber\\
&=& \sum_{m=1,2}F_{lm}(\tilde{\lambda}) K_m + \sum_{m=1,2}
\left[G_m(\tilde{\lambda}) T \delta_{lm} + a H_{lm}(\tilde{\lambda})
K_m -\frac{a}{\pi}F'_{lm}(\tilde{\lambda})K_m \right]
\sqrt{\tilde{\ell}} + O(\tilde{\ell}) \label{eq:SIFsTilde}
\end{eqnarray}
(where second order terms with respect to $a$ have been
disregarded). This expression is of the form
\begin{equation}
K'_l(\ell)=\left[K'_l(\tilde{\ell})\right]_{a=0}^{\pi\tilde{\lambda}}+a\sum_{m}\tilde{H}_{lm}(\tilde{\lambda})K_m
\sqrt{\tilde{\ell}} + O(\tilde{\ell}),
\label{eq:Hlm}
\end{equation}
where $\left[K'_l(\tilde{\ell})\right]_{a=0}^{\pi\tilde{\lambda}}$
is the $l$'th SIF at the tip of a straight extension of length
$\tilde{\ell}$ in the direction $\tilde{\lambda}$, and the functions
$\tilde{H}_{lm}$ are defined by $\tilde{H}_{lm}=H_{lm}-F'_{lm}/\pi$,
so we can invert it to get
\begin{equation}
H_{lm}(\lambda)=\tilde{H}_{lm}(\lambda)+\frac{F_{lm}'(\lambda)}{\pi}.
\label{eq:Hlm2}
\end{equation}
The expression (\ref{eq:Hlm}) for the SIFs, which is exact to the
first order with respect to $a$, is precisely of the form which will
result from the perturbative analysis. It will therefore be easy to
identify the functions $\tilde{H}_{lm}$, and the $H_{lm}$'s will
follow using Eq.~(\ref{eq:Hlm2}).

\subsubsection{Perturbative analysis, integral equations}

The equations of the problem with curved extensions in the physical $Z$-plane take the same
form as in the case of straight extensions, where again $\Phi$ and
$\Psi$ denote the real Muskhelishvili potentials. We associate some
new potentials $\Phi^S$ and $\Psi^S$ with $\Phi$ and $\Psi$ through
analytic continuation by shifting the curved extension into
the fictitious straight one. $\Phi^S$ and $\Psi^S$ have discontinuities across the fictitious straight extensions, while $\Phi$ and $\Psi$ have discontinuities across the actual curved extensions. The values of $\Phi^S$ and $\Psi^S$ at each side of the fictitious straight
extension are analytic continuations of the values taken by $\Phi$ and $\Psi$  on both sides of the curved extensions. If $Z=Z_F$ represents the points of the straight fictitious extension ($Z_F^{\pm}$ will represent points on opposite sides of the extensions), and if
$Z_F+\eta_u(Z_F)$ represents the points of the curved extension of the upper ($u$) branch, then the original and shifted potentials are related through:
\bqe
\Phi \left(Z_F^{\pm} + \eta_{u}(Z_F)\right)  \simeq \Phi^{S}(Z_F^{\pm})+\eta_u(Z_F)\Phi'^{S}(Z_F^{\pm}) +O\left(\eta^2\right) \, .
\label{rela}
\eqe
Also, $\Phi^S$ and $\Psi^S$ are expanded in powers of $\eta$:
 \begin{eqnarray}
 \Phi^S & = & \Phi_0+\Phi_1+O(\eta^2) \label{Sexp} \, , \\
 \Psi^S & = & \Psi_0+\Psi_1+O(\eta^2) \, .
 \end{eqnarray}
 Combining Eqs.~(\ref{rela}) and (\ref{Sexp}):
\bqe
\Phi \left(Z_F^{\pm} + \eta_{u}(Z_F)\right)  \simeq
\Phi_0(Z_F^{\pm})+\Phi_1(Z_F^{\pm}) +\Phi_0'(Z_F^{\pm}) \eta_u(Z_F) \, .
\eqe
Following the perturbative analysis of \citep{amle} one obtains the following forms for Eq.~(\ref{PP}) written in $z$-space to orders $O(1)$ and $O(\eta)$ respectively:
\begin{eqnarray}
Const & = & \phi_0(z)+\frac{\omega(z)}{\overline{\omega'(z)}}
\overline{\phi'_0(z)}+\overline{\psi_0(z)} \label{O0} \, , \\
Const' & = & \phi_1(z) +
\omega(z) \frac{\overline{\phi_1'(z)}}{\overline{\omega'(z)}}+\overline{\psi_1(z)}
+\left[\frac{\phi_0'(z)}{\omega'(z)}+\frac{\overline{\phi_0'(z)}}{\overline{\omega'(z)}}\right]\eta(z) \nonumber \\
&+& \left[\omega(z)\frac{\overline{\phi_0''(z)}}{\left(\overline{\omega'(z)}\right)^2}-\omega(z)
\frac{\overline{\omega''(z)}}{\left(\overline{\omega'(z)}\right)^3}\overline{\phi_0'(z)}+\frac{\overline{\psi_0'(z)}}{\overline{\omega'(z)}}\right]
\overline{\eta(z)} \; , \label{eqpe}
\end{eqnarray}
for $z \in \tilde {\cal C}$, with $\tilde {\cal C}$ the unit circle in $z$ space corresponding to the fictitious star shaped crack ($\phi_0(z)=\Phi_0(Z)$, etc.). Notice that Eq. (\ref{O0}) is equivalent to Eq. (\ref{pp}), and that Eq. (\ref{eqle}) follows from the latter.
Applying the previously mentioned Lemma (in section
\ref{section:straight}, right after Eq.~(\ref{Qz})) to Eq.~(\ref{eqle}) (to its right hand side, and considered for the fictitious
crack), one obtains:
\begin{equation}
\psi_0(z) =
(\Gamma+\Gamma')C' z-\Gamma \overline{C'}/z +Q_*(z)\phi_0'(z)-
\chi_{1*}^{(0)}(z)-\chi_{2*}^{(0)}(z) \; ,
\label{foz}
\end{equation}
where
\begin{equation}
\chi_{1,2*}^{(0)}(z) = \frac{(1-e^{\pm
2i\tilde{\lambda}\pi})}{2\pi i}z \int_{\tilde {\cal C}_{1,2}} d\eta
\frac{Q_*(\eta)\phi_0'(\eta)}{\eta(z-\eta)} \; ,
\end{equation}
and with $Q_*(\eta)$ given by Eq.~(\ref{qast}) (with fictitious parameters). The poles of $Q_*(\eta)$
at $e^{i\tilde \beta}$ and $e^{-i\tilde \beta}$ are displaced into $\Omega^+$ (for
$Q(\eta)$ they are at $\Omega^-$).

The problem now is to solve Eq.~(\ref{eqpe}). Analysis of the
potentials close to the singular points follows closely the one of \citep{amle}. In order to apply the Lemma to Eq.~(\ref{eqpe}), one defines the functions:
\begin{equation}
\chi_{1,2*}^{(1)}(z)
= \frac{(1-e^{\mp 2i\tilde{\lambda}\pi})}{2\pi i} \int_{\tilde{\cal C}_{1,2}}
d\eta \frac{Q(\eta)\overline{\phi_1'(\eta)}}{(\eta-z)} \; ,
\end{equation}
i.e. by Plemelj's formula,
they do satisfy:
\begin{equation}
\chi_{1,2*}^{(1)}(z^+)-\chi_{1,2*}^{(1)}(z^-)=(1-e^{\mp 2i\tilde{\lambda}\pi})Q(z)\overline{\phi_1'(z)} \, ,
\end{equation}
if $z$ belongs to $\tilde{\cal C}_{1,2}$ respectively, and zero otherwise.
Also one defines:
\begin{eqnarray}
\phi_{1,2}^{0(1)}(z) &\equiv& \frac{1}{2\pi
i} \int_{\tilde{\cal C}_{1,2}}\frac{dt}{(t-z)}
\left[\frac{\phi_0'(t)}{\omega'(t)}+\frac{\overline{\phi_0'(t)}}{\overline{\omega'(t)}}\right]\eta(t) \nonumber \\
&+& \left[\frac{\omega(t)
\overline{\phi_0''(t)}}{(\overline{\omega'(t)})^2}-\frac{\omega(t)\overline{\omega''(t)} \overline{\phi_0'(t)}}{(\overline{\omega'(t)})^3}
+\frac{\overline{\psi_0'(t)}}{\overline{\omega'(t)}}\right]\overline{\eta(t)} \label{f231} \, .
\end{eqnarray}
Plemelj's formula can be applied in an analogous way to this
equation. Applying these previous equations, Eq.~(\ref{eqpe}) can
be written as:
\begin{eqnarray}
& & \phi_1(z)-\chi_1^{(1)}(z^-)-\chi_2^{(1)}(z^-)-\phi_1^{0(1)}(z^-)-\phi_2^{0(1)}(z^-) \nonumber \\
& & = \overline{Q_*(z)\phi_1'(z)-\chi_{1*}^{(1)}(z^-)-\chi_{2*}^{(1)}(z^-)
-\phi_{1*}^{0(1)}(z^-)-\phi_{2*}^{0(1)}(z^-)} + Const.
\end{eqnarray}
for $z \in \tilde {\cal C}$. The Lemma implies that: \bqe
\phi_1(z)-\chi_1^{(1)}(z^-)-\chi_2^{(1)}(z^-)-\phi_1^{0(1)}(z^-)-
\phi_2^{0(1)}(z^-)=Const. \eqe Differentiating this previous
equation, one obtains:
\begin{equation}
\phi_1'(z)=\phi_1^{0(1)'}(z)+\phi_2^{0(1)'}(z)+\tilde{\mathcal{L}} \phi_{1}'(z)
\label{f1p},
\end{equation}
where $\tilde{\mathcal{L}}$ is the
operator of Eq.~(\ref{Lop}), with $\lambda \rightarrow
\tilde{\lambda}$.

The steps in order to obtain the stress intensity factors at the
tips of the curved extensions are the following:

\begin{itemize}
\item i) Solve for $\phi_0'(z)$ from the following equation:
\begin{equation}
\phi_0'(z)=\phi'^{(0)}_0(z)+\tilde{\mathcal{L}}(\phi_0'(z)) \; ,
\label{eqfio}
\end{equation}
(that follows from from Eq.~(\ref{O0})) with
\begin{equation}
 \phi'^{(0)}_0(z) = \Gamma C + (\Gamma+\overline{\Gamma'})\overline{C}/z^2.
 \label{fiop}
\end{equation}
(This step was practically done in section \ref{section:straight}
with $\lambda$ and $\ell$ instead of $\tilde\lambda$ and
$\tilde\ell$.)
 \item ii) Evaluate $\psi_0(z)$ from Eq.~(\ref{foz}).
 \item iii) Calculate $\phi_{1,2}^{0(1)}(z)$ from Eq.~(\ref{f231}).
 \item iv) Solve the integral equation (\ref{f1p}) for $\phi_1'(z)$.
 \item v) Get the stress intensity factors using Andersson's formula (\ref{And})
 with $\phi'$ replaced by $\phi'^S=\phi_0'+\phi_1'$.
 \end{itemize}

\subsubsection{Expansion to order $1/2$ in the extension length}

We will now expand the preceding equation up to order
$\sqrt{\tilde\ell}$, or equivalently up to order $\tilde\alpha$ ( $\tilde\alpha$ is given by Eq.~(\ref{alfa}) with $\lambda \rightarrow
\tilde\lambda$ and $\ell \rightarrow \tilde \ell$). This will
yield the functions $H_{lm}$. For that purpose we perform a change of
variable, $z=e^{i\alpha \zeta}$, and the following expansions of
functions:
\begin{eqnarray}
\phi_0'(z) & = & e^{-i\tilde\alpha \zeta}[\sqrt{L}U_0(\zeta)+
\tilde\alpha L V_0(\zeta)+O(\tilde\alpha^2)] \, , \\
\psi_0'(z) & = & e^{-i \tilde\alpha \zeta}[\sqrt{L} X_0(\zeta) +O(\tilde\alpha)] \, ,  \\
\phi_1'(z) & = & e^{-i \tilde\alpha \zeta}[\sqrt{L} U_1(\zeta) +\tilde\alpha L V_1(\zeta) +O(\tilde\alpha^2)] \, .
\label{expp}
\end{eqnarray}
Expansion of the integral equation (\ref{eqfio}) for $\phi_0'(z)$ up
to order $\tilde\alpha$ leads to the following integral equations for
$U_0$ and $V_0$:
\begin{eqnarray}
U_0(\zeta) & = & U^{(0)}_0(\zeta)+\tilde{\mathcal{A}} U_0(\zeta) \label{u0} \, ,\\
V_0(\zeta) & = & V^{(0)}_0(\zeta)+\tilde{\mathcal{A}} V_0(\zeta) \label{v0} \;,
\end{eqnarray}
where $\tilde{\mathcal{A}}$ is the operator $\mathcal{A}$ of Eq.~(\ref{Afm}) with $\lambda \rightarrow \tilde{\lambda}$, and $ U^{(0)}_0(\zeta)= U^{(0)}_0$,  $ V^{(0)}_0(\zeta)= V^{(0)}_0$ are equivalent to those of Eqs.~(\ref{eq:U})-(\ref{eq:V0}) ($\lambda \rightarrow \tilde{\lambda}$).
Therefore, based on the results of section \ref{section:straight},
we can consider these as known functions from now on.

The expressions for $U'_0$ and $X_0$ will also be needed here. The
first one is obtained simply by differentiating Eq. (\ref{u0}) once
with respect to $\zeta$. In addition, we decompose $U_0(\zeta)$ into
$U_{0,j}(\zeta)$ ($j=1,2$) as is done in Eq. (\ref{decomp:U}), i.e. $U_0(\zeta)=(K_1 U_{0,1}(\zeta)-i K_2 U_{0,2}(\zeta))/\sqrt{8\pi}$, so we get for
$U'_{0,j}(\zeta)$
\begin{equation}
U'_{0,j}(\zeta) = \pm \frac{(1-e^{-2i\tilde{\lambda}\pi})}{2\pi i}
\int_{C_1^+} d\eta \frac{\eta
(\eta^2-1)\overline{U_{0,j}}(\eta)}{(\eta^2-\tilde{\lambda})(\eta-\zeta)^3}
\pm  \frac{(1-e^{2i\tilde{\lambda}\pi})}{2\pi i} \int_{C_2^+}
d\eta \frac{\eta
(\eta^2-1)\overline{U_{0,j}}(\eta)}{(\eta^2-\tilde{\lambda})(\eta-\zeta)^3}
\; ,
\end{equation}
where the upper and lower cases on $\pm$ correspond
to $j=1,2$ respectively, and $C_{1,2}^+$ are convenient deformations of the paths $\eta
\in [-1,0]$ and $\eta \in [0,1]$ into $\Omega^+$, respectively
(in this way the poles at $\zeta=\pm
\sqrt{\tilde{\lambda}}-i\epsilon$ are avoided) as before - see Fig
\ref{contours}.

In order to obtain $X_0(\zeta)$ we differentiate and expand the
expression for $\psi_0'(z)$ following from Eq.~(\ref{foz})
\begin{eqnarray}
X_0(\zeta) & = & \frac{K_1+iK_2}{\sqrt{8\pi}}-\frac{(\zeta^4+(1-3\tilde{\lambda})\zeta^2+\tilde{\lambda})}{2(\zeta^2-\tilde{\lambda})^2}U_0(\zeta) - \frac{\zeta(\zeta^2-1)}{2(\zeta^2-\tilde{\lambda})} U_0'(\zeta) \nonumber \\
& & -\frac{(1-e^{-2i\tilde{\lambda}\pi})}{4\pi i}\int_{-1}^0dh \frac{h(h^2-1)U_0(h)}{(h^2-\tilde{\lambda})(h-\zeta)^2}-\frac{(1-e^{2i\tilde{\lambda}\pi})}{4\pi i}\int_0^1 dh \frac{h(h^2-1)U_0(h)}{(h^2-\tilde{\lambda})(h-\zeta)^2} \; ,
\end{eqnarray}
where now the poles are located at $\zeta=\pm
\sqrt{\tilde{\lambda}}+i\epsilon$ (see the comment after Eq.~(\ref{qast})).
Therefore, in this expression the integrals over $[-1,0]$ and
$[0,1]$ are deformed away from the poles into contours in the lower
half-plane. However, using the lower semi-circles $C_1^- \cup C_2^-$
as integration contours is not wise because then we encounter
singularities in the integrand, as we are interested in
evaluating $X_0(\zeta)$ on $C_1^- \cup C_2^-$. It is wiser to use the
two lower semi-ellipses $\Gamma_{1,2}^-$ defined by $\eta = \left( \mp 1 -
( \cos \theta + i \sin \theta/2) \right)/2, \;\;
0<\theta<\pi$ (see Fig. \ref{contours}). In order to take
advantage of the resolution of $U_{0,j}(\zeta)$ ($j=1,2$) done in
the section \ref{section:straight}, we decompose $X_0$ as in Eq.~(\ref{decomp:U})
\begin{equation}
X_0(\zeta) = \frac{1}{\sqrt {8\pi}}(K_1X_{0,1}(\zeta) -
i K_2 X_{0,2}(\zeta) ) \, . \label{decomp:X0}
\end{equation}
And we get the following two expressions ($j=1,2$)
\begin{eqnarray}
 X_{0,j} \left( \zeta  \right) &=& \left( { - 1} \right)^{j + 1}  - \frac{{(\zeta ^4  + \left( {1 - 3\tilde{\lambda} )} \right)\zeta ^2  + \tilde{\lambda }}}{{2\left( {\zeta ^2  - \tilde{\lambda} } \right)^2 }}U_{0,j} \left( \zeta  \right) - \frac{{\zeta \left( {\zeta ^2  - 1} \right)}}{{2\left( {\zeta ^2  -\tilde{ \lambda} } \right)}}U'_{0,j} (\zeta)  \nonumber\\
&-& \frac{{1 - e^{ - 2i\tilde{\lambda} \pi } }}{{4i\pi }}\int_{\Gamma _1^ -  }^{} {\frac{{\eta \left( {\eta^2  - 1} \right)}}{{\eta^2  - \lambda }}\frac{{U_{0,j} \left( \eta  \right)d\eta }}{{\left( {\eta  - \zeta } \right)^2 }}}  - \frac{{1 - e^{2i\lambda \pi } }}{{4i\pi }}\int_{\Gamma _2^ -  }^{} {\frac{{\eta \left( {\eta^2  - 1} \right)}}{{\eta^2  - \lambda }}\frac{{U_{0,j} \left( \eta  \right)d\eta }}{{\left( {\eta  - \zeta } \right)^2 }}}.
\end{eqnarray}

We will now expand the integral equation (\ref{f1p}) for $\phi'_1$ in
powers of $\tilde{\alpha}$. The first step is to get an expansion of
$\omega(z)$ and its derivatives:
\begin{eqnarray}
\omega(z)   & = & -\frac{L}{4}\tilde\alpha^2(\zeta^2-1)\left(\frac{\zeta}{\zeta-1}\right)^{\tilde{\lambda}}\left(\frac{\zeta}{\zeta+1}\right)^{\tilde{\lambda}} \, ,\\
\omega'(z)  & = &  \frac{i\tilde\alpha L}{2\zeta}(\zeta^2-\tilde{\lambda})\left(\frac{\zeta}{\zeta-1}\right)^{\tilde{\lambda}}\left(\frac{\zeta}{\zeta+1}\right)^{\tilde{\lambda}} \, , \\
\omega''(z) & = & \frac{L}{2} \left(\frac{\zeta}{\zeta-1}\right)^{\tilde{\lambda}}\left(\frac{\zeta}{\zeta+1}\right)^{\tilde{\lambda}} \frac{(\zeta^4-(1+\tilde{\lambda})\zeta^2+2\tilde{\lambda}^2-\tilde{\lambda})}{\zeta^2(\zeta^2-1)} \, .
\end{eqnarray}
An expression for the gap $\eta_u(z)$ between the fictitious
and curved upper branch is the following:
\begin{equation}
\eta_u(z) =- ia e^{i\pi \tilde{\lambda}}|Z|(\sqrt{\tilde{\ell}}-\sqrt{|Z|}) \, .
\end{equation}
Since
\begin{eqnarray}
\left|\omega(z)\right| & = & |Z| \simeq \frac{L}{4} \tilde{\alpha}^2
\left|\zeta^2-1\right|
\left|\frac{\zeta}{\zeta+1}\right|^{\tilde{\lambda}}\left|\frac{\zeta}{\zeta-1}\right|^{\tilde{\lambda}} \, , \\
\sqrt{\tilde{\ell}} &\simeq&
\sqrt{\frac{(1-\tilde{\lambda})L}{4}}\left(\frac{\tilde{\lambda}}{1-\tilde{\lambda}}\right)^{\tilde{\lambda}/2}\tilde{\alpha} \, ,
\end{eqnarray}
it follows that
\begin{equation}
\eta_u(z) = ia e^{i\pi \tilde{\lambda}}\left(\frac{L}{4}\right)^{3/2}\tilde{\alpha}^3\left|\zeta^2-1\right|\left|\frac{\zeta}{\zeta+1}\right|^{\tilde{\lambda}}\left|\frac{\zeta}{\zeta-1}\right|^{\tilde{\lambda}} G(\zeta) \; ,
\end{equation}
with
\begin{equation}
G(\zeta) \equiv
\sqrt{\left|\zeta^2-1\right|}\left|\frac{\zeta}{\zeta+1}\right|^{\tilde{\lambda}/2}\left|\frac{\zeta}{\zeta-1}\right|^{\tilde{\lambda}/2}
-\sqrt{1-\tilde{\lambda}}\left(\frac{\tilde{\lambda}}{1-\tilde{\lambda}}\right)^{\tilde{\lambda}/2} \, .
\end{equation}

All the necessary ingredients for expanding $\phi_2^{0(1)'}(z)$ and
the integral equation (\ref{f1p}) are now available. Starting from
Eq.~(\ref{f231}), one gets the following expansion
\begin{eqnarray}
\phi_2^{0(1)'}(z) =
&-& \frac{a L\tilde{\alpha}}{8\pi} \int_0^1 dh
\frac{h|h^2-1|G(h)}{(h^2-\tilde{\lambda})(h-\zeta)^2}
\left\{ U_0(h)+\overline{X_0(h)} \right. \nonumber \\
&+& \left. \frac{e^{i2\tilde{\lambda}\pi}}{2(h^2-\tilde{\lambda})}\left[h(h^2-1)\overline{U_0'(h)}-
\frac{(3h^4-(1+5\tilde{\lambda})h^2+4\tilde{\lambda}^2-\tilde{\lambda})}{(h^2-\tilde{\lambda})}\overline{U_0(h)}\right]
\right\} \; , \label{fi201}
\end{eqnarray}
with a pole at
$\zeta=\sqrt{\tilde{\lambda}}-i\epsilon$. Similarly, at the lower
branch
\bqe
\eta_l(z) = -iae^{-i\pi \tilde{\lambda}}\left(\frac{L}{4}\right)^{3/2}\tilde{\alpha}^3\left|\zeta^2-1\right|\left|\frac{\zeta}{\zeta+1}\right|^{\tilde{\lambda}}\left|\frac{\zeta}{\zeta-1}\right|^{\tilde{\lambda}} G(\zeta) \; ,
\eqe
and then
\begin{eqnarray}
\phi_1^{0(1)'}(z) &=& \frac{a L\tilde{\alpha}}{8\pi} \int_{-1}^0 dh
\frac{h|h^2-1|G(h)}{(h^2-\tilde{\lambda})(h-\zeta)^2}
\left\{ U_0(h)+\overline{X_0(h)} \right. \nonumber \\
&& + \left.\frac{e^{-i2\tilde{\lambda}\pi}}{2(h^2-\tilde{\lambda})}\left[h(h^2-1)\overline{U_0'(h)}-
\frac{(3h^4-(1+5\tilde{\lambda})h^2+4\tilde{\lambda}^2-\tilde{\lambda})}{(h^2-\tilde{\lambda})}\overline{U_0(h)}\right]
\right\} \; , \label{fi101}
\end{eqnarray}
with a pole at $\zeta=-\sqrt{\tilde{\lambda}}-i\epsilon$. In order
to calculate more efficiently the integrals in Eqs.~(\ref{fi201}),
(\ref{fi101}), we would need to deform the contours (as done many
times before) into $C_{1,2}^+$. For that purpose the function
$G(\zeta)$ may be continued analytically into disks that encircle
the segments $\zeta \in [-1,0]$ and $\zeta \in [0,1]$. These
continuations are respectively:
\begin{eqnarray}
G(\zeta) & = & e^{-i
\pi(1+\tilde{\lambda})/2}\zeta^{\tilde{\lambda}}(\zeta+1)^{(1-\tilde{\lambda})/2}(\zeta-1)^{(1-\tilde{\lambda})/2}
-\sqrt{1-\tilde{\lambda}}\left(\frac{\tilde{\lambda}}{1-\tilde{\lambda}}\right)^{\tilde{\lambda}/2} \, , \\
G(\zeta) & = & e^{-i
\pi(1-\tilde{\lambda})/2}\zeta^{\tilde{\lambda}}(\zeta+1)^{(1-\tilde{\lambda})/2}(\zeta-1)^{(1-\tilde{\lambda})/2}
-\sqrt{1-\tilde{\lambda}}\left(\frac{\tilde{\lambda}}{1-\tilde{\lambda}}\right)^{\tilde{\lambda}/2} \, .
\end{eqnarray}

The following step is to solve Eq.~(\ref{f1p}) for $\phi_1'(z)$.
Replacing the expansion for $\phi_1'(z)$ in Eq.~(\ref{expp}) into
Eq.~(\ref{f1p}), one gets that $\phi_{1,2}^{0(1)'}(z)$ are of order
$\tilde{\alpha}$. As a result the equation for $U_1$ (analogous to Eqs.~(\ref{u0})-(\ref{v0}) for $U_0$ and $V_0$) becomes
$U_1(\zeta)=\tilde{\mathcal{A}} U_1(\zeta)$ (where here too
$\tilde{\mathcal{A}}$ is the same operator as in Eq.~(\ref{Afm}),
but with $\lambda \rightarrow \tilde{\lambda}$). Due to the
contracting nature of $\tilde{\mathcal{A}}$ (see Appendix A) this
implies that $U_1=0$. Expanding Eq.~(\ref{f1p}) to order $\tilde{\alpha}$,
using the previous expressions for $\phi_{1,2}^{0(1)'}(z)$ (Eqs.~(\ref{fi101}), (\ref{fi201})), we get
\begin{equation}
V_1(\zeta)=V_1^{(0)}(\zeta)+\tilde{\mathcal{A}}V_1(\zeta) \, ,
\label{v1ex}
\end{equation}
with
\begin{eqnarray}
V_1^{(0)}(\zeta) & = & \frac{a}{8\pi} \left\{ \int_{-1}^0 dh
\frac{h|h^2-1|G(h)}{(h^2-\tilde{\lambda})(h-\zeta)^2}
\left\{ U_0(h)+\overline{X_0(h)} \right. \right.\nonumber \\
& & + \left.\frac{e^{-i2\tilde{\lambda}\pi}}{2(h^2-\tilde{\lambda})}\left[h(h^2-1)\overline{U_0'(h)}-
\frac{(3h^4-(1+5\tilde{\lambda})h^2+4\tilde{\lambda}^2-\tilde{\lambda})}{(h^2-\tilde{\lambda})}\overline{U_0(h)}\right] \right\} \nonumber \\
& & - \int_0^1 dh \frac{h|h^2-1|G(h)}{(h^2-\tilde{\lambda})(h-\zeta)^2}
\left\{ U_0(h)+\overline{X_0(h)} \right. \nonumber \\
& & + \left. \left. \frac{e^{i2\tilde{\lambda}\pi}}{2(h^2-\tilde{\lambda})}\left[h(h^2-1)\overline{U_0'(h)}-
\frac{(3h^4-(1+5\tilde{\lambda})h^2+4\tilde{\lambda}^2-\tilde{\lambda})}{(h^2-\tilde{\lambda})}\overline{U_0(h)}\right]
\right\} \right\} \, .
\end{eqnarray}
By further decomposing $V_1(\zeta)$ into
\begin{equation}
V_1(\zeta) = \frac{a}{(8\pi)^{3/2}} \left(K_1 V_{1,1} -
i K_2 V_{1,2} \right) \, , \label{decomp:V1}
\end{equation}
we get:
\begin{eqnarray}
V_{1,j}^{(0)}(\zeta) & = &  \int_{-1}^0 dh
\frac{h|h^2-1|G(h)}{(h^2-\tilde{\lambda})(h-\zeta)^2}
\left\{ U_{0,j}(h) \pm \overline{X_{0,j}(h)} \right. \nonumber \\
& & \pm \left. \frac{e^{-i2\tilde{\lambda}\pi}}{2(h^2-\tilde{\lambda})}\left[h(h^2-1)\overline{U_{0,j}'(h)}-
\frac{(3h^4-(1+5\tilde{\lambda})h^2+4\tilde{\lambda}^2-\tilde{\lambda})}{(h^2-\tilde{\lambda})}\overline{U_{0,j}(h)}\right] \right\} \nonumber \\
& & - \int_0^1 dh \frac{h|h^2-1|G(h)}{(h^2-\tilde{\lambda})(h-\zeta)^2}
\left\{ U_{0,j}(h) \pm \overline{X_{0,j}(h)} \nonumber \right.\\
& & \left.\pm \frac{e^{i2\tilde{\lambda}\pi}}{2(h^2-\tilde{\lambda})}\left[h(h^2-1)\overline{U_{0,j}'(h)}-
\frac{(3h^4-(1+5\tilde{\lambda})h^2+4\tilde{\lambda}^2-\tilde{\lambda})}{(h^2-\tilde{\lambda})}\overline{U_{0,j}(h)}\right]
\right\} \; ,
\label{eq:V10}
\end{eqnarray}
(the upper and lower cases of $\pm$ correspond to $j=1,2$
respectively) where previous decompositions for $U_0(\zeta)$ and
$X_0(\zeta)$ (given by Eqs.~(\ref{decomp:U}),(\ref{decomp:X0})) have
been used. Then, Eq.~(\ref{v1ex}) leads to the following equations
to be solved:
\begin{eqnarray}
V_{1,1}(\zeta) & = & V_{1,1}^{(0)}(\zeta)+\tilde\mathcal{A} V_{1,1}(\zeta) \nonumber \, , \\
V_{1,2}(\zeta) & = & V_{1,2}^{(0)}(\zeta)-\tilde\mathcal{A} V_{1,2}(\zeta) \; ,
\end{eqnarray}
which are solved as in the previous section by iterations as follows:
\begin{eqnarray}
V_{1,1}(\zeta) & = & \sum_{n=0}^{\infty} \tilde\mathcal{A}^n V_{1,1}^{(0)}(\zeta) \, , \\
V_{1,2}(\zeta) & = & \sum_{n=0}^{\infty} (-1)^n \tilde\mathcal{A}^n V_{1,2}^{(0)}(\zeta) \, .
\end{eqnarray}

In order to apply Andersson's formula (\ref{And}) at the upper
branch tip, one uses
\begin{eqnarray}
\delta & \simeq &  \pi \lambda +\frac{3}{2}a\sqrt{\tilde{\ell}} \, , \\
\omega''(e^{i\tilde\beta}) & = & \frac{\omega(e^{i\tilde\beta})
(e^{i\tilde\beta}+1)(e^{i\tilde\beta}-e^{-i\tilde\beta})}
{e^{i\tilde\beta}(e^{i\tilde\beta}-e^{i\tilde\alpha})(e^{i\tilde\beta}-e^{-i\tilde\alpha})(e^{i\tilde\beta}-1)} \simeq L e^{i\tilde{\lambda}
\pi}\left(1-2i\sqrt{\tilde{\lambda}}
\tilde\alpha\right)\left(\frac{\tilde{\lambda}}{1-\tilde{\lambda}}\right)^{\tilde{\lambda}} \, .
\end{eqnarray}
Also, since $\tilde\beta \simeq \sqrt{\tilde{\lambda}} \tilde\alpha$, and
$\tilde{\lambda}=\lambda+a\sqrt{\ell}/\pi$ we get
\begin{equation}
K'_1(\tilde{\ell})-iK'_2(\tilde{\ell}) =
2\sqrt{\pi}\left[U\left(\sqrt{\tilde{\lambda}}\right)+\sqrt{L}\tilde\alpha
V\left(
\sqrt{\tilde{\lambda}}\right)\right]\left(\frac{1-\tilde{\lambda}}{\tilde{\lambda}}\right)^{\tilde{\lambda}/2}
e^{-i\pi \tilde{\lambda}} e^{-ia\sqrt{\tilde{\ell}}/4} \, .
\end{equation}
Then, for the curved crack case one can write (Eq.~(\ref{alfa})):
\begin{eqnarray}
&& K'_1(\tilde{\ell}) - iK'_2(\tilde{\ell}) =
\left[{K'_1(\tilde{\ell})-iK'_2(\tilde{\ell})}\right]_{a=0}^{\pi \tilde{\lambda}} \nonumber \\
&& + 2\sqrt{\pi}e^{-i\tilde{\lambda}\pi}\left(\frac{1-\tilde{\lambda}}{\tilde{\lambda}}\right)^{\tilde{\lambda}/2}\left[-i\frac{a}{4}U_0\left(\sqrt{\tilde{\lambda}}\right)+
\frac{2}{\sqrt{1-\tilde{\lambda}}}\left(\frac{1-\tilde{\lambda}}{\tilde{\lambda}}\right)^{\tilde{\lambda}/2}V_1\left(\sqrt{\tilde{\lambda}}\right)
\right]\sqrt{\tilde{\ell}} \, . \label{k12}
\end{eqnarray}
Using the decompositions given by Eqs.~(\ref{decomp:U}),
(\ref{decomp:X0}) and (\ref{decomp:V1}), the last expression for the
stress intensity factors becomes:
\begin{eqnarray}
&& K'_1(\tilde{\ell})-iK'_2(\tilde{\ell}) = \left[K'_1(\tilde{\ell})-iK'_2(\tilde{\ell})\right]_{a=0}^{\pi \tilde{\lambda}} \nonumber \\
&& + \frac{a}{4\sqrt2}\left(\frac{1-\tilde{\lambda}}{\tilde{\lambda}}\right)^{\tilde{\lambda}/2}e^{-i\tilde{\lambda}\pi}
\left\{K_1\left[-iU_{0,1}\left(\sqrt{\tilde{\lambda}}\right)+\frac{1}{\pi}
\sqrt{\frac{1}{1-\tilde{\lambda}}}\left(\frac{1-\tilde{\lambda}}{\tilde{\lambda}}\right)^{\tilde{\lambda}/2}V_{1,1}\left(\sqrt{\tilde{\lambda}}\right)
\right] \right. \nonumber \\
&& - \left. K_2\left[U_{0,2}\left(\sqrt{\tilde{\lambda}}\right)+
\frac{i}{\pi}\sqrt{\frac{1}{1-\tilde{\lambda}}}\left(\frac{1-\tilde{\lambda}}{\tilde{\lambda}}\right)^{\tilde{\lambda}/2}V_{1,2}\left(\sqrt{\tilde{\lambda}}\right)
\right] \right\} \sqrt{\tilde{\ell}} \, .
\label{k12e}
\end{eqnarray}
The $\tilde{H}_{lm}$'s are extracted from the defining expression
Eq.~(\ref{eq:Hlm}) together with the previous result. Notice that
extracting the $\tilde{H}_{lm}$'s from the last equation involves
evaluating the functions $V_{1,j}(\zeta)$ at $\sqrt{\tilde\lambda}$ (on the
real axis). This involves, as explained in  \citep{amle}, crossing of
the pole at $\eta=\sqrt{\tilde\lambda}$ for $V_{1,j}^{(0)}(\zeta)$. Appropriate account
of this difficulty results in
that if one just replaces $\zeta=\sqrt{\tilde\lambda}$ in expression
(\ref{eq:V10}) the contribution from the $U$'s should be doubled. Thus, we get:
\begin{eqnarray}
\tilde{H}_{11}(\tilde{\lambda}) & = &
\frac{-1}{4\sqrt2}\left(\frac{1-\tilde{\lambda}}{\tilde{\lambda}}\right)^{\tilde{\lambda}/2}
\Re \left\{ e^{-i\tilde{\lambda}\pi} \left[
2iU_{0,1}\left(\sqrt{\tilde{\lambda}}\right)-
\frac{1}{\pi}\sqrt{\frac{1}{1-\tilde{\lambda}}}\left(\frac{1-\tilde{\lambda}}{\tilde{\lambda}}\right)^{\tilde{\lambda}/2}V_{1,1}\left(\sqrt{\tilde{\lambda}}\right) \right] \right\}  \, , \\
\tilde{H}_{12}(\tilde{\lambda}) & = &
\frac{-1}{4\sqrt2}\left(\frac{1-\tilde{\lambda}}{\tilde{\lambda}}\right)^{\tilde{\lambda}/2}
\Re \left\{ e^{-i\tilde{\lambda}\pi} \left[
2U_{0,2}\left(\sqrt{\tilde{\lambda}}\right)+
\frac{i}{\pi} \sqrt{\frac{1}{1-\tilde{\lambda}}}\left(\frac{1-\tilde{\lambda}}{\tilde{\lambda}}\right)^{\tilde{\lambda}/2}V_{1,2}\left(\sqrt{\tilde{\lambda}}\right) \right] \right\} \, ,\\
\tilde{H}_{21}(\tilde{\lambda}) & = &
\frac{1}{4\sqrt2}\left(\frac{1-\tilde{\lambda}}{\tilde{\lambda}}\right)^{\tilde{\lambda}/2}
\Im \left\{ e^{-i\tilde{\lambda}\pi} \left[
2iU_{0,1}\left(\sqrt{\tilde{\lambda}}\right)-
\frac{1}{\pi}\sqrt{\frac{1}{1-\tilde{\lambda}}}\left(\frac{1-\tilde{\lambda}}{\tilde{\lambda}}\right)^{\tilde{\lambda}/2}V_{1,1}\left(\sqrt{\tilde{\lambda}}\right) \right] \right\} \, ,\\
\tilde{H}_{22}(\tilde{\lambda}) & = &
\frac{1}{4\sqrt2}\left(\frac{1-\tilde{\lambda}}{\tilde{\lambda}}\right)^{\tilde{\lambda}/2}
\Im \left\{ e^{-i\tilde{\lambda}\pi} \left[
2U_{0,2}\left(\sqrt{\tilde{\lambda}}\right)+
\frac{i}{\pi}\sqrt{\frac{1}{1-\tilde{\lambda}}}\left(\frac{1-\tilde{\lambda}}{\tilde{\lambda}}\right)^{\tilde{\lambda}/2}V_{1,2}\left(\sqrt{\tilde{\lambda}}\right)
\right] \right\} \;  .
\end{eqnarray}
Finally, Eq.~(\ref{eq:Hlm2}) is used in order to extract the
$H_{lm}$'s from the $\tilde{H}_{lm}$'s. Results following from a
numerical resolution of the problem were presented above in Fig.
\ref{fig:Hlm}. In that figure, we superimposed the results obtained
in \citep{amle} for the case of a kinked crack (with the same
angle) for the sake of comparison (actually in \citep{amle} only
results up to $\lambda=80^o$ are presented, so we extended the
results obtained there to the whole range $\lambda \in [0,1]$ in
order to allow for a comparison between the two cases).

\section{Elastostatic analysis of a crack in a strip geometry}
\label{section:T}

\begin{figure}[ht]
\centerline{\includegraphics[width=8cm]{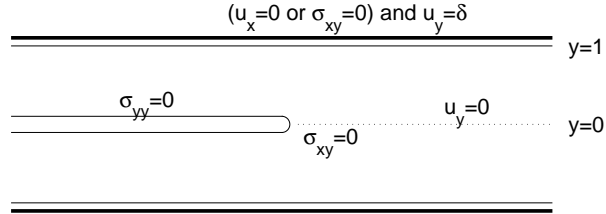}}
\caption{A semi-infinite crack in a strip of unit half-width. The
top and bottom edges of the strip are fixed at height $\pm\delta$ above
their initial positions, and either these edges are clamped (first problem) or free to slide (second problem).}
\label{fig:Tstresss}
\end{figure}

In this section we calculate the ratio $T/K_1$ for a static semi-infinite crack in a strip of unit half-width in an isotropic elastic medium described by Lam\'e coefficientes $\lambda$, $\mu$ (see Fig.~\ref{fig:Tstresss}). The system is loaded by imposing normal displacements $\pm\delta$ to the top and bottom edges of the strip. We will consider both clamped ($u_x=0$, first problem) and shear free ($\sigma_{xy}=0$, second problem) boundary conditions. In both cases, the symmetry of the problem allows to focus on the upper half plane only. Also, we subtract the solution of the unbroken strip, which corresponds to a state of uniform stress. This contribution will be added when needed. The boundary conditions corresponding to the first problem are
\begin{eqnarray}
u_x(x,1) &=&u_y(x,1)=\sigma_{xy}(x,0)=0 \;,
\label{eq:bcA1}\\
\sigma_{yy}(x,0)&=& -\sigma_\infty  \, ,\qquad  |x|<0\;,
\label{eq:bcA4}\\
u_y(x,0) & = &0\,,\qquad |x|>0 \, , \label{eq:bcA5}
\end{eqnarray}
while the second problem has the following boundary conditions
\begin{eqnarray}
\sigma_{xy}(x,1) &=&u_y(x,1)=\sigma_{xy}(x,0)=0 \;,
\label{eq:bcB1}\\
\sigma_{yy}(x,0)&=& -\sigma_\infty  \,,\qquad  |x|<0\;,
\label{eq:bcB4}\\
u_y(x,0) & = &0\,,\qquad |x|>0 \label{eq:bcB5} \, .
\end{eqnarray}
Here $\sigma_\infty$ can be determined from the solution of the unbroken strip and its value is different for each problem. Notice that for the second problem, any addition of a constant stress
$\sigma_{xx}(x \rightarrow \pm \infty,y) \equiv T_{\infty}$ is consistent
with the boundary conditions, and thus with the formulation of this
problem. In both cases we have
\begin{equation}
u_y(x \rightarrow -\infty, y \rightarrow 0^+)=\delta
\label{eq:delta} \, ,
\end{equation}

Introducing Fourier transforms in the $x$-coordinates as $y(x)=\frac{1}{2\pi}\int^\infty_{-\infty} Y(k) e^{-\imath kx} dk$, the elastostatic equilibrium equations can be solved without difficulties. Moreover, applying the boundary conditions (\ref{eq:bcA1}) and (\ref{eq:bcB1}) of the first and second problem respectively, allows to reduce the determination of the stress field to the problem of solving the following equation with the appropriate boundary conditions:
\begin{equation}
(\lambda+2\mu)F(k) \equiv -\frac{\Sigma_{yy}(k)}{U_y(k)} \label{eq:Fk1} \, ,
\end{equation}
where
\begin{eqnarray}
\Sigma_{yy}(k) &\equiv& \Sigma_{yy}(k,0)=\int^\infty_{-\infty} \sigma_{yy}(x,0) e^{\imath kx} dk \label{eq:Syy2} \, ,\\
U_y(k)  &\equiv& U_y(k,0)=\int^\infty_{-\infty} u_{yy}(x,0) e^{\imath kx} dk \label{eq:uy2} \, ,
\end{eqnarray}
and $F(k)$ is a known function that is obtained in a closed form for each problem. As will be explicitly shown below, the behavior of $F(k)$ at large and small $k$'s is given by
\begin{eqnarray}
F(k) &\simeq& f_0 \, , \qquad |k|\ll1 \label{eq:Fk3} \, ,\\
F(k) &\simeq& f_\infty |k| \, , \qquad |k|\gg1 \label{eq:Fk4} \, .
\end{eqnarray}

The piecewise boundary conditions (\ref{eq:bcA4}),(\ref{eq:bcA5}) and (\ref{eq:bcB4}),(\ref{eq:bcB5}) suggest the use of the Wiener-Hopf decomposition method. Let us introduce
\begin{eqnarray}
\Sigma_{yy}^+(k) &=& \int_0^\infty \sigma_{yy}(x,0) e^{\imath kx} dx\;,\\
\Sigma_{yy}^-(k) &=& \int_{-\infty}^0 \sigma_{yy}(x,0) e^{\imath kx} dx\;,
\end{eqnarray}
with $U_y^+(k)$ and $U_y^-(k)$ defined similarly. Notice that
$\Sigma_{yy}^-(k)$ ($U_y^-(k)$) is analytic for $\Im k<0$ and
$\Sigma_{yy}^+(k)$ ($U_y^+(k)$) is analytic for $\Im k>0$. Using~(\ref{eq:bcA4}),(\ref{eq:bcA5}) and (\ref{eq:bcB4}),(\ref{eq:bcB5}) together with Eq.~(\ref{eq:Fk1}) one thus obtains
\begin{equation}
-(\lambda+2\mu)F(k)U_y^-(k)=\Sigma_{yy}^+(k) - \frac
{\sigma_\infty}{\imath k} \label{eq:Fk2} \, .
\end{equation}
Let us suppose that one can write
\begin{equation}
F(k) = \frac{F^-(k)}{F^+(k)} \, , \label{eq:Fk5}
\end{equation}
where $F^-(k)$ has neither zeros nor poles for $\Im k<0$
and $F^+(k)$ has neither zeros nor poles for $\Im k>0$.
Then Eq.~(\ref {eq:Fk2}) is rewritten as
\begin{equation}
-(\lambda+2\mu)kF^-(k)U_y^-(k)=k\Sigma_{yy}^+(k)F^+(k) +\imath
\sigma_\infty F^+(k) \label{eq:Fk6} \, .
\end{equation}
The left-hand side of Eq.~(\ref{eq:Fk6}) is analytic in the lower half plane, while its right-hand side is analytic in the upper half plane, and both sides coincide on the real axis. By the theorem of analytic continuation, and in order to retrieve the square root behavior of the stress field at the crack tip, both sides must equal a constant. This constant can be fixed by examining the behavior of the expression for $k \rightarrow 0$. In fact Eq. (\ref{eq:delta}) implies that
\begin{equation}
U^-_y(k) \simeq  \frac{\delta}{\imath k}\;,\qquad  k \rightarrow 0\;.
\label{eq:smallk}
\end{equation}
Equations (\ref{eq:Fk6}) and (\ref{eq:smallk}) then imply
\begin{equation}
\Sigma_{yy}(k)\equiv\Sigma_{yy}^+ - \frac
{\sigma_\infty}{\imath k}=-\frac{\sigma_0}{\imath k }\frac{F^-(0)}{F^+(k)}\;, \label{eq:Syysol}
\end{equation}
where
\begin{equation}
\sigma_0 = (\lambda+2\mu)\delta \label{eq:sigma0} \, .
\end{equation}
Now, since we are interested in the $T$-stress we look at the difference
\begin{equation}
\sigma_{xx}(x,0^+)-\sigma_{yy}(x,0^+) =\frac{1}{2\pi}
\int_{-\infty}^\infty G(k) \Sigma_{yy}(k) e^{-\imath k x} dk
\label{eq:T1} \, ,
\end{equation}
where $G(k)=-1+\Sigma_{xx}(k)/\Sigma_{yy}(k)$ will be simply a given function for each problem. We will also need the stress intensity factor (SIF) $K_1$ in each case, for which the expression is given by
\begin{equation}
K_1 = \lim_{|k| \rightarrow \infty}\left[\sqrt{-2\imath
k}\sigma_{yy}(k)\right] =\sigma_0\sqrt{2 f_0f_\infty} \label{eq:K1} \, .
\end{equation}
Before examining each problem separately, the method of decomposition of $F(k)$ as $F^-(k)/F^+(k)$ which is necessary for the Wiener-Hopf method will be presented. To do this, let us define the function
\begin{equation}
H(k)=\frac{F(k)}{\sqrt{k^2f^2_\infty + f_0^2}} \label{eq:g} \, .
\end{equation}
where $f_0$ and $f_\infty$ are chosen to get the two limits $|k| \rightarrow 0,\pm\infty$ right
for real $k$. The function $H(k)$ is a bounded, even function, which tends to $1$
for $|k| \rightarrow 0,\pm\infty$. Therefore, we can approximate this function using the
method of Pad\'e \citep{Baker}. This method was first used in the context of fracture by \citep{Bouch03}.

A {\it{Pad\'e approximant}}, is that rational function of a specified order whose power series expansion agrees with a given power series to the highest possible order. In the present case, the Pad\'e approximation of $g(k)$ will be of the form
\begin{equation}
H(k)\simeq\frac{P_{2N}(k)}{Q_{2N}(k)} \label{eq:pade1} \, ,
\end{equation}
where $P_{2N}(k)$ and $Q_{2N}(k)$ are polynomials, which must be
even, and $2N$ is the order of the approximation (in practice we use
$2N=30$ which gives accuracy better than $10^{-5}$ for the desired
quantities). Then, we find all the complex roots
$\ell_1,...,\ell_{2N}$ and $\lambda_1,...,\lambda_{2N}$ of
$P_{2N}(k)$ and $Q_{2N}(k)$ respectively (actually it is even a
simpler problem to find the $N$ roots of the polynomials
$P_{2N}(\sqrt{x})$ and $Q_{2N}(\sqrt{x})$, denoted by $r_1,...,r_N$
and $\rho_1,...,\rho_N$ respectively, from which we recover
$\ell_{2n-1},\ell_{2n} = \pm \sqrt{r_n}$ and
$\lambda_{2n-1},\lambda_{2n} = \pm \sqrt{\rho_n}$), and so we get
the factorization
\begin{equation}
H(k)\simeq\frac{(1-\frac{k^2}{r_1}) \cdots
(1-\frac{k^2}{r_N})}{(1-\frac{k^2}{\rho_1}) \cdots
(1-\frac{k^2}{\rho_N})} \label{eq:pade2} \, .
\end{equation}
The Wiener-Hopf decomposition of a function of this form may be
carried out by inspection:
\begin{equation}
F^+(k)= \frac{1} {\sqrt{f_0-\imath k f_\infty}}
\prod_{n=1}^{N}\left(\frac{1-\frac{\imath k}{\sqrt{r_n}}}
{1-\frac{\imath k}{\sqrt{\rho_n}}}\right) \label{eq:pade3} \, .
\end{equation}
This approximation schemes converges when taking larger and larger
$N$'s, and so the choice of $N$ is a matter of the desired accuracy.
The advantage of this approach over expanding in Chebyshev
polynomials, as done for example in \citep{LiuMarder}, is
twofold. First, by using the same number of series coefficients a
better approximation for $g(k)$ is obtained. Second, when
factorizing the expansion in Chebyshev polynomials to order $2N$ one
has to find roots of a polynomial of order $2N$ while for the
comparable Pad\'e approximation one has to factorize two polynomials
of order $N$ which is simpler.

\subsection{Solution of the first problem}

\begin{figure}[ht]
\centerline{\includegraphics[width=8cm]{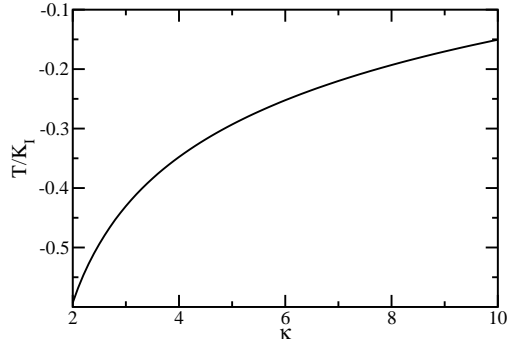}}
\caption{The dimensionless ratio $T/K_1$ as a function of $\kappa$ for the clamped boundary condition. Notice that for $\kappa=3$, $T/K_1=-0.43$, or in dimensioned quantities $T/K_1=-0.5/\sqrt{W}$, where $W$ is the width of the sample.} \label{fig:TPr1}
\end{figure}

For the problem defined by the boundary conditions (\ref{eq:bcA1})-(\ref{eq:bcA5}), the function $F(k)$ is defined by
\begin{equation}
F(k)=\frac{2}{\kappa^2} \frac{k\left[(\kappa^2+1)+2(\kappa-1)^2k^2 +
(\kappa^2-1)\cosh(2k) \right]}{(\kappa+1)\sinh(2k)-2k(\kappa-1)} \, ,
\label{eq:FkPr1}
\end{equation}
so that $f_0 = 1$ and $f_\infty = 2(\kappa-1)/\kappa^2$. Also, the function $G(k)$ is given by
\begin{equation}
G(k)=-4\frac{(\kappa-1)^2k^2+\kappa}{(\kappa^2+1)+2(\kappa-1)^2k^2 +
(\kappa^2-1)\cosh(2k)} \label{eq:GkPr1} \, ,
\end{equation}
By solving the problem of an unbroken strip with the same boundary
conditions one gets
\begin{equation}
\sigma_\infty = \sigma_0 =(\lambda+2\mu)\delta \, .
\label{eq:sigmainfPr1}
\end{equation}
Now, using the decomposition given by Eqs.~(\ref{eq:Syysol}) and (\ref{eq:pade3}), one has
\begin{equation}
\Sigma_{yy}(k)=-\frac{\sigma_0}{\imath k}\sqrt{1-\imath k
f_\infty}\prod_{n=1}^{N}\left(\frac{1-\frac{\imath
k}{\sqrt{\rho_n}}} {1-\frac{\imath k}{\sqrt{r_n}}}\right) \, .
\label{eq:SyyPrl}
\end{equation}
The stress intensity factor given by Eq.~(\ref{eq:K1}) yields
\begin{equation}
K_1 = 2\sigma_0\sqrt{\kappa-1}/\kappa\;.
\label{eq:K11}
\end{equation}
The $T$-stress is related to the asymptotic value of $(\sigma_{xx}-\sigma_{yy})$ at
$y=0$ for $x \rightarrow 0$ with adding the solution of the unbroken strip
\begin{equation}
T = \lim_{x \rightarrow
0}\left[\sigma_{xx}(x,0^+)-\sigma_{yy}(x,0^+)\right] -
\frac{2\mu}{\lambda+2\mu} \sigma_0 =\frac{1}{2\pi}
\int_{-\infty}^\infty G(k) \Sigma_{yy}(k) dk - \frac{2}{\kappa}
\sigma_0 \label{eq:T11} \, .
\end{equation}
In Fig. \ref{fig:TPr1}, the ratio $T/K_1$ is given as a function of $\kappa$. Notice that $T$ is always negative for this problem.

\subsection{Solution of the second problem}

\begin{figure}[ht]
\centerline{\includegraphics[width=8cm]{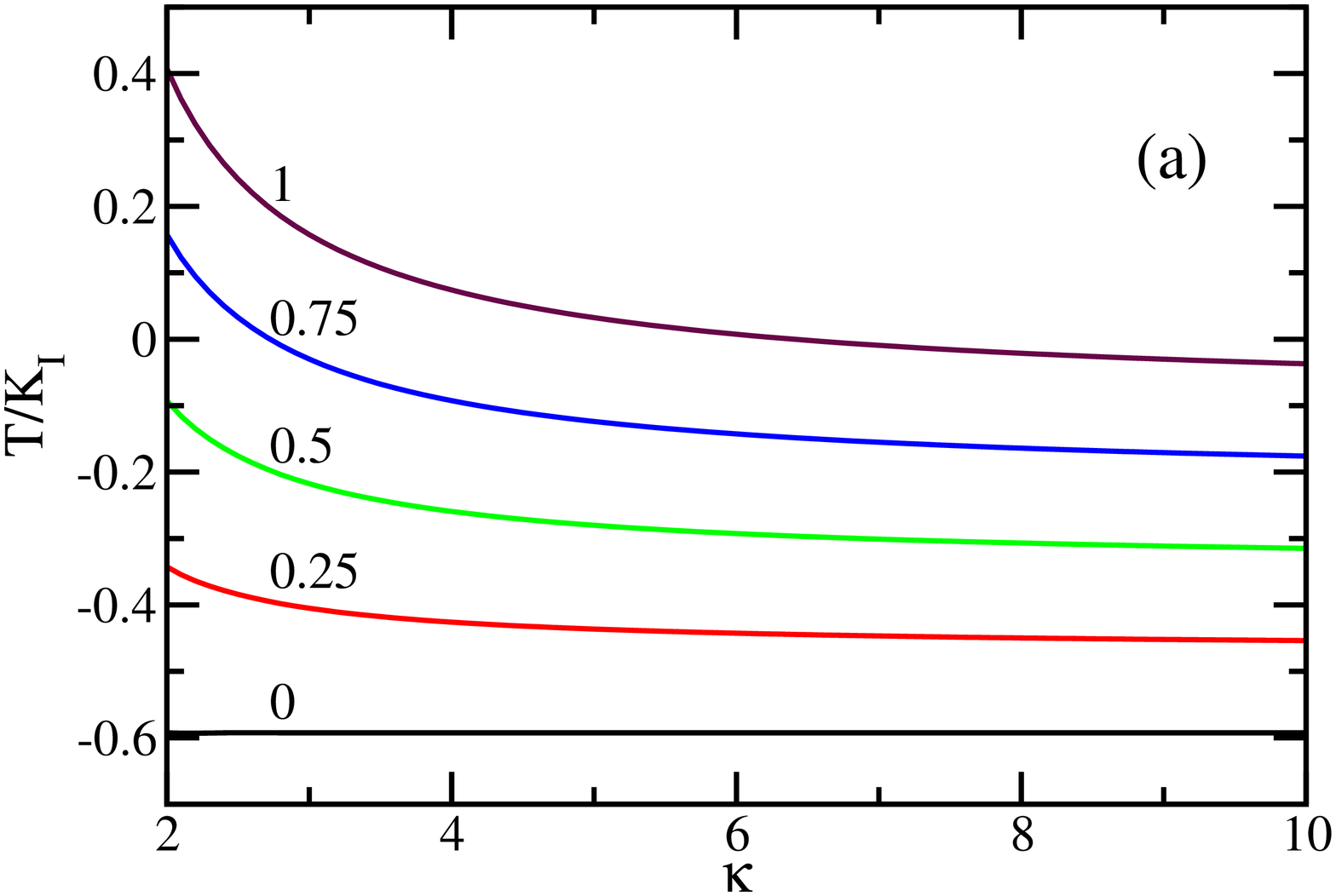}
\includegraphics[width=8cm]{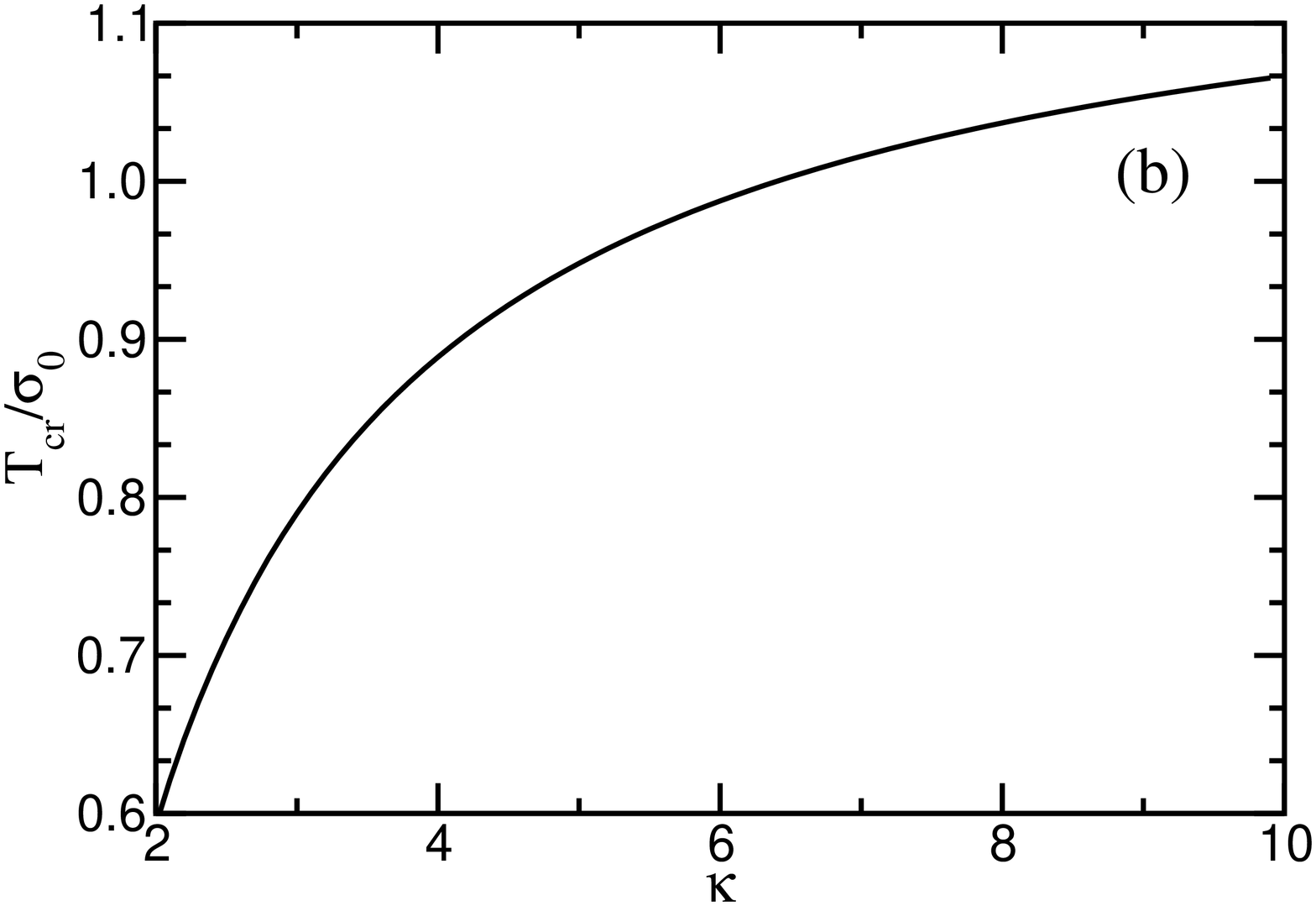}}
\caption{(a) The dimensionless  ratio $T/K_1$ as a function of $\kappa$ for the shear free boundary condition. The different curves correspond to different values of $T_\infty/\sigma_0$.
(b) $T_{cr}/\sigma_0$ for various values of $\kappa$.} \label{fig:TPr2}
\end{figure}

For the problem defined by the boundary conditions (\ref{eq:bcB1})-(\ref{eq:bcB5}), the function $F(k)$ is
\begin{equation}
F(k)=\frac{(\kappa-1)}{\kappa^2} \frac{k(2k+\sinh(2k))}{\sinh^2(k)} \, ,
\label{eq:FkPr2}
\end{equation}
so that $f_0 = 4(\kappa-1)/\kappa^2$ and $f_\infty = 2(\kappa-1)/\kappa^2$. Moreover, one has
\begin{equation}
G(k)=\frac{-4k}{2k+\sinh(2k)} \label{eq:GkPr2} \, ,
\end{equation}
and
\begin{equation}
\sigma_\infty = \sigma_0 \left[
1+ \frac{\kappa-2}{\kappa}  \left( \frac{T_\infty}{\sigma_0}
-\frac{\kappa-2}{\kappa} \right) \right] \label{eq:sigmainfPr2} \, ,
\end{equation}
where $T_{\infty} \equiv \sigma_{xx}(\infty)$.
Now, using again the decomposition given by Equations (\ref{eq:Syysol}) and
(\ref{eq:pade3}), one gets
\begin{equation}
\Sigma_{yy}(k)=-\frac{f_0\sigma_0}{\imath k}\sqrt{1-\imath k
/2}\prod_{n=1}^{N}\left(\frac{1-\frac{\imath k}{\sqrt{\rho_n}}}
{1-\frac{\imath k}{\sqrt{r_n}}}\right) \, . \label{eq:SyyPr2}
\end{equation}
In this case, the stress intensity factor is given by
\begin{equation}
K_1 = \frac{4(\kappa-1)}{\kappa^2}\sigma_0 \label{eq:K12}\;,
\end{equation}
and
\begin{equation}
T = \frac{1}{2\pi} \int_{-\infty}^\infty
G(k) \Sigma_{yy}(k) dk - \frac{2}{\kappa}\sigma_0 \left[ \frac{2(\kappa-1)}{\kappa} -
\frac{T_\infty}{\sigma_0} \right]
\label{eq:T12} \, .
\end{equation}
Fig.~\ref{fig:TPr2}a shows the behavior of $T/K_1$ as a function of $\kappa$ for various
values of $T_\infty$. For every value of $\kappa$ there exists a critical value of
$T_\infty$, denoted by $T_{cr}$, at which the value of $T$
changes sign (see Fig. \ref{fig:TPr2}b).

Since a change in the sign of $T$ implies a transition from a stable
crack growth ($T<0$) to an unstable one ($T>0$), a destabilization/stabilization of the
growth process may be induced by tuning the value of $T_\infty$. Interestingly, this feature is not shared by the first
problem (clamped edges) and is a particular property of the shear free
boundary conditions.

\section*{Acknowledgements}

This work was supported by EEC PatForm Marie Curie action (E.K.), and by an international cooperation CNRS-Conicyt. Laboratoire de Physique Statistique is associated with Universities Paris VI and Paris VII.

\newpage
\section*{Appendix A: Proof that $\mathcal{A}$ is contracting}

Take the space of functions which are defined and continuous on
$C_{1,2}^-$, such that the norm defined by
\begin{equation}
\left\| f \right\| = \mathop {Max}\limits_{\zeta  \in C_{1,2}^-}
\left| {\left( {\zeta ^2  - 1} \right)\zeta f\left( \zeta  \right)}
\right|
\end{equation}
is finite. Here, we intend to prove that the operator $\mathcal{A}$ (given by Eq.~(\ref{Afm}))
is contracting in this space, i.e. that there exists a constant $c$
smaller than $1$ such that
\begin{equation}
\left\| \mathcal{A}{f} \right\| \le c\left\| f \right\|
\label{App:bound}
\end{equation}
for every function in that space. Rewriting $\mathcal{A}$ by shifting the
contours to $C_{1,2}^-$ we get
\begin{eqnarray}
\mathcal{A} f\left( \zeta  \right) &=& \frac{{1 - e^{ - 2i\lambda \pi } }}{{8\pi }}\int_0^\pi  {\frac{{\left( { - {\textstyle{{1 + e^{ - i\theta } } \over 2}}} \right)\left[ {\left( { - {\textstyle{{1 + e^{ - i\theta } } \over 2}}} \right)^2  - 1} \right]}}{{\left( { - {\textstyle{{1 + e^{ - i\theta } } \over 2}}} \right)^2  - \lambda }}\frac{{\overline {f\left( { - {\textstyle{{1 + e^{ - i\theta } } \over 2}}} \right)} e^{ - i\theta } d\theta }}{{\left[ {\left( { - {\textstyle{{1 + e^{ - i\theta } } \over 2}}} \right) - \zeta } \right]^2 }}}  +  \nonumber\\
&+&  \frac{{1 - e^{2i\lambda \pi } }}{{8\pi }}\int_0^\pi
{\frac{{\left( {{\textstyle{{1 - e^{ - i\theta } } \over 2}}}
\right)\left[ {\left( {{\textstyle{{1 - e^{ - i\theta } } \over 2}}}
\right)^2  - 1} \right]}}{{\left( {{\textstyle{{1 - e^{ - i\theta }
} \over 2}}} \right)^2  - \lambda }}\frac{{\overline {f\left( { -
{\textstyle{{1 + e^{ - i\theta } } \over 2}}} \right)} e^{ - i\theta
} d\theta }}{{\left[ {\left( {{\textstyle{{1 - e^{ - i\theta } }
\over 2}}} \right) - \zeta } \right]^2 }}}.
\end{eqnarray}
Now using $\left| {\frac{{1 - e^{ \pm 2i\lambda \pi } }}{{8\pi }}}
\right| = \frac{{\sin \pi \lambda }}{{4\pi }}$, $\left| {\frac{{1 -
e^{ - i\theta } }}{2}} \right| = \sin \frac{\theta }{2}$, $\left|
{\frac{{ - 1 - e^{ - i\theta } }}{2}} \right| = \cos \frac{\theta
}{2}$, $\left| {\left( {\frac{{1 - e^{ - i\theta } }}{2}} \right)^2
- 1} \right| = \sqrt {\frac{{5 - 3\cos \theta }}{2}} \cos
\frac{\theta }{2}$ and $\left| {\left( {\frac{{ - 1 - e^{ - i\theta
} }}{2}} \right)^2  - 1} \right| = \sqrt {\frac{{5 + 3\cos \theta
}}{2}} \sin \frac{\theta }{2}$, we get the following estimate
\begin{eqnarray}
\left\| {\mathcal{A}f} \right\| &\le& \left\| f \right\| \cdot
\mathop {Max}\limits_{\zeta  \in C_{1,2}^ -  } \left\{ {\frac{{\sin
\pi \lambda }}{{4\pi }}\left| {\left( {\zeta ^2  - 1} \right)\zeta }
\right| }\right.\nonumber\\
&\cdot& \left.{ \int_0^\pi  {d\theta \left| {{\textstyle{1 \over
{\left[ {\left( { - {\textstyle{{1 + e^{ - i\theta } } \over 2}}}
\right) - \zeta } \right]^2 \left[ {\left( { - {\textstyle{{1 + e^{
- i\theta } } \over 2}}} \right)^2  - \lambda } \right]}}} +
{\textstyle{1 \over {\left[ {\left( {{\textstyle{{1 - e^{ - i\theta
} } \over 2}}} \right) - \zeta } \right]^2 \left[ {\left(
{{\textstyle{{1 - e^{ - i\theta } } \over 2}}} \right)^2  - \lambda
} \right]}}}} \right|} } \right\}.
\end{eqnarray}
We focus on the left branch $C_1^-$, where $\zeta=-\frac{{1 +
e^{i\gamma } }}{2}$ ($\gamma  \in \left[ {0,\pi } \right]$). There
we get $\left| (\zeta^2-1)\zeta \right|= \sqrt {\frac{{5 + 3\cos
\gamma }}{8}} \sin \gamma$. As for the integral in the last
inequality, we are not able to bound it analytically, but it is not
difficult to show numerically that it obtains its maximum at $\lambda
\rightarrow 1$ and $\gamma \rightarrow \pi$ where it behaves like
$\frac{sin(\pi\lambda)}{\pi (1-\lambda)}$. And so for every
$\lambda<1$ we get that $\mathop {Max}\limits_{\zeta  \in C_1^-} \{
\cdots \}<1$. Similarly, for the right branch $C_2^-$, where
$\zeta=\frac{{1 - e^{i\gamma } }}{2}$ ($\gamma  \in \left[ {0,\pi }
\right]$) we get exactly the same bound, since the only difference
with respect to $C_1^-$ is $\gamma \rightarrow \pi-\gamma$.
Therefore we get what we need, i.e. there exists a constant $c<1$
such that Eq.~(\ref{App:bound}) is obeyed.

\end{document}